\documentclass[%
 reprint,
 superscriptaddress,
 amsmath,amssymb,
 aps,
 floatfix,
 footinbib
]{revtex4-1}

\usepackage{amsmath}
\usepackage{graphicx}
\usepackage{dcolumn}
\newcolumntype{d}[1]{D{.}{.}{#1}}

\usepackage{bm}
\usepackage{epstopdf}
\usepackage{graphicx}
\usepackage{stmaryrd}
\usepackage{scalerel,amssymb}
\usepackage{tikz}
\usepackage{pgf}
\usepackage{xcolor,cancel}

\definecolor{lightblue}{RGB}{0,170,255}

\newcommand{\appropto}{\mathrel{\vcenter{
			\offinterlineskip\halign{\hfil$##$\cr
				\propto\cr\noalign{\kern2pt}\sim\cr\noalign{\kern-2pt}}}}}

\usepackage[normalem]{ulem}
\usepackage{color}
\usepackage{xr}

\begin{document}
	
	\title{Non-linear spin filter for non-magnetic materials at zero magnetic field}
	
	\author{E. Marcellina}
	\altaffiliation{Present address: School of Physical and Mathematical Sciences, Nanyang Technological University, 21 Nanyang Link, Singapore 637371. Email: emarcellina@ntu.edu.sg}
	\affiliation{School of Physics, The University of New South Wales, Sydney, Australia}
			
	\author{A. Srinivasan}
	\affiliation{School of Physics, The University of New South Wales, Sydney, Australia}
	
	\author{F. Nichele}
	\affiliation{IBM Research - Zurich,	S\"{a}umerstrasse 4, 8803 R\"{u}schlikon, Switzerland}
	
	\author{P. Stano}
	\affiliation{RIKEN Center for Emergent Matter Science (CEMS), Wako, Saitama 351-0198, Japan}
	\affiliation{Department of Applied Physics, School of Engineering, University of Tokyo, 7-3-1 Hongo, Bunkyo-ku, Tokyo 113-8656, Japan}
	\affiliation{Institute of Physics, Slovak Academy of Sciences, 845 11 Bratislava, Slovakia}
	
	\author{D. A. Ritchie}
	\affiliation{Cavendish Laboratory, University of Cambridge, J. J. Thomson Avenue, Cambridge CB3 0HE, United Kingdom}
	
	\author{I. Farrer}
	\affiliation{Cavendish Laboratory, University of Cambridge, J. J. Thomson Avenue, Cambridge CB3 0HE, United Kingdom}
	
	\author{Dimitrie Culcer}
	\affiliation{School of Physics, The University of New South Wales, Sydney, Australia}
	
	\author{A. R. Hamilton}
	\affiliation{School of Physics, The University of New South Wales, Sydney, Australia}
	
\date{\today}
	
\begin{abstract}
The ability to convert spin accumulation to charge currents is essential for applications in spintronics. In semiconductors, spin-to-charge conversion is typically achieved using the inverse spin Hall effect or using a large magnetic field. Here we demonstrate a general method that exploits the non-linear interactions between spin and charge currents to perform all-electrical, rapid and non-invasive detection of spin accumulation without the need for a magnetic field. We demonstrate the operation of this technique with ballistic GaAs holes as a model system with strong spin-orbit coupling, in which a quantum point contact provides the non-linear energy filter. This approach is generally applicable to electron and hole systems with strong spin orbit coupling.
\end{abstract}
	
\maketitle

\textit{Introduction}. Spintronics is a technology that uses the spin degree of freedom to manipulate information \cite{Wolf2001,Zutic2004}.  A key challenge in spintronics is the generation and detection of spin accumulation \cite{Han2018}. In semiconductors, spin accumulation is typically generated by optical excitations \cite{Stern2006,Chang2007,Matsuzaka2009,Ando2010,Wunderlich2010} or the intrinsic spin Hall effect \cite{Bruene2010,Balakrishnan2013a,Choi2015, Bottegoni2017}, whilst spin-to-charge conversion (i.e. spin accumulation translating into a charge current or voltage) is achieved through the inverse spin Hall effect \cite{Bruene2010,Balakrishnan2013a,Choi2015}. However, generating/detecting spin accumulation optically or via the spin Hall effect-inverse spin Hall effect pair is challenging for strongly spin-orbit coupled mesoscopic systems with short spin relaxation time and spin diffusion lengths.

Here we adapt the concept of a spin filter, i.e. a device that separates spin species based on their energies, for detecting spin accumulation in strongly spin-orbit coupled mesoscopic systems. The first spin filter was developed by Stern and Gerlach who used an inhomogenous magnetic field to spatially separate electrons with different spins \cite{SternGerlach1922}. Spin filters have also been realized in the solid state using spin-dependent transport in mesoscopic devices \cite{Chesi2011,Nichele2015}. These techniques allow a spin current to be converted into a charge current, which is then detected as a voltage signal that depends on the applied magnetic field. Unfortunately, these linear techniques require a large magnetic field, which is impractical and can change the spin signal being probed. In this work, we demonstrate a non-linear technique that requires no magnetic field, and allows fast detection of spin accumulation. 

We use GaAs holes as a model system for strongly spin-orbit coupled systems with short spin relaxation time ($< 100$ fs) \cite{Hilton2002} and  spin diffusion length much shorter than the typical device dimensions ($\sim 100 - 1000$ nm, see also Sec.~S2 of the Supplementary Material). Semiconductor holes have recently attracted great interest in semiconductor spintronics due to their exceptionally strong spin-orbit interaction \cite{Akhgar2016, Wang2016, Srinivasan2017, Marcellina2018, Hendrickx2018, Watzinger2018, Crippa2018, Hendrickx2018, Hendrickx2020}. The spin accumulation in strongly spin-orbit coupled ballistic mesoscopic systems is generated as follows. In mesoscopic systems with strong spin-orbit interaction, charge currents are generally accompanied by spin currents \cite{Bardarson2007,Krich2008,Adagideli2010,Stano2011,Ramos2018}. When the spin-orbit length is much shorter than both the device dimensions and mean free path, the spin precesses around randomly oriented spin-orbit fields throughout the sample region, giving rise to spin currents with a non-zero average \cite{Adagideli2010}. Consequently, different spin species can have different chemical potentials, which give rise to a net spin accumulation whose amount and distribution depend on the sample geometry as well as the strength and form of the spin-orbit interaction. The spin accumulation adjacent to the energy barrier can then be detected through a voltage signal containing contributions linear and non-linear in spin accumulation.

This paper is laid out as follows. We first demonstrate spin-to-charge conversion in the linear regime using an in-plane magnetic field. We then show spin-to-charge conversion in the non-linear regime and confirm that it works even at zero magnetic field, so that is all-electrical and works much faster than linear spin-to-charge conversion. Our method can be generalized for any strongly spin-orbit coupled material such as GaSb, InAs, transition metal dichalcogenides, as well as topological insulators, since non-linear spin-to-charge conversion only requires a finite spin accumulation, regardless of the spin orientation, and an energy barrier. Furthermore, the rapidness of non-linear spin-to-charge conversion enables detection of spin orientation with radio-frequency techniques  down to 1 ns \cite{Taskinen2008}. 

\begin{figure}[t]
	\begin{center}
		\includegraphics[scale = 1.9,trim={0cm 2cm 1.5cm 0cm}]{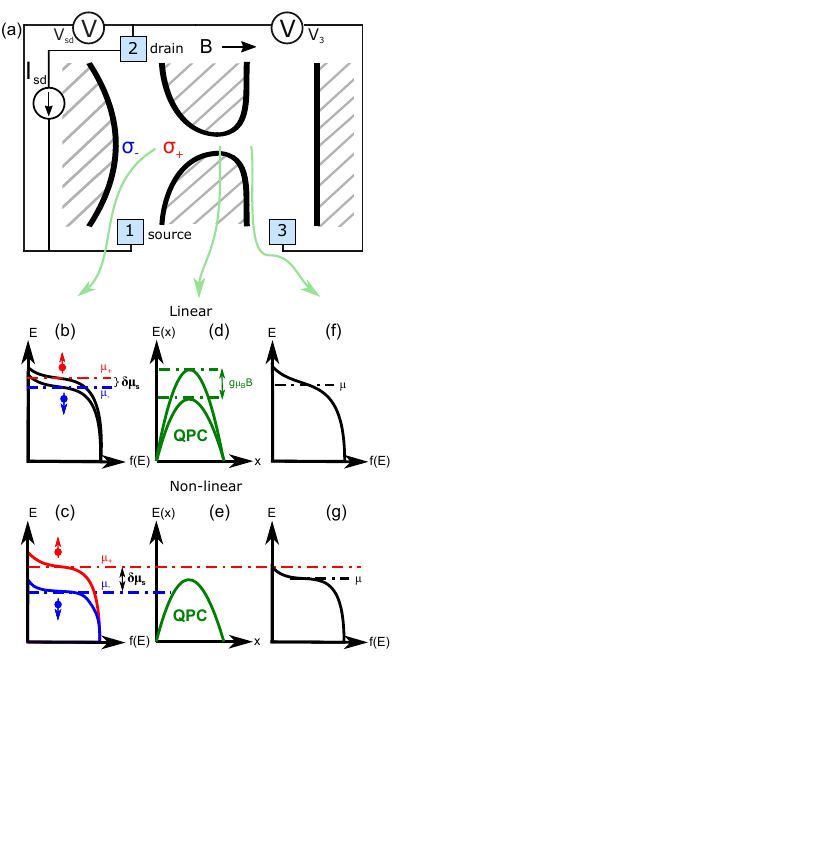}
		\caption{(a) A schematic of the experimental setup. A current $I_{\mathrm{sd}}$ flows between terminals 1 and 2, resulting in a voltage difference $V_{\mathrm{sd}}$ across the drive channel. (b), (c) Near the quantum point contact (QPC), opposite spin orientations $\sigma_+$ and $\sigma_-$ accumulate on opposite sides of the drive channel. The QPC acts as an energy filter: Spin-to-charge conversion occurs due to the difference in the transmission probability through the QPC between each spin species.  In the linear regime (d), this difference arises from a different kinetic energy caused by, for instance, a Zeeman interaction. (e) In the non-linear regime, the different local chemical potentials for $\sigma_+$ and $\sigma_-$ give rise to different transmission probabilities. In both (f) linear and (g) non-linear cases, spin-polarized holes accumulate after passing through the QPC, resulting in a voltage $V_3$ between terminals 2 and 3. Note that the schematics in (a)-(g) are not to scale.}
		\label{fig:spin_accumulation}
	\end{center}
\end{figure}	

\textit{Experimental concept}. We use a three-terminal geometry with a quantum point contact (QPC) as an energy-selective barrier  (Fig.~\ref{fig:spin_accumulation}a). Passing a current $I_\mathrm{sd}$ in the drive channel between terminals 1 and 2 results in a voltage difference $V_{\mathrm{sd}}$ and a net non-equilibrium spin accumulation $\delta{\mu}_s$: Spins with orientation $\sigma_+$ have a higher chemical potential (of $\delta{\mu}_s$)  than  $\sigma_-$ (Figs.~\ref{fig:spin_accumulation}b and c). The kink in the drive channel helps direct the spin accumulation towards the QPC \cite{Benter2013}. Spin-to-charge conversion occurs if one spin species has a higher transmission probability $T(E)$ through the QPC than the other. In the linear regime, the difference in the transmission probability originates from the difference in the hole's kinetic energy due to an in-plane Zeeman interaction (Fig.~\ref{fig:spin_accumulation}d). However, in the non-linear regime, the energy dependence of the transmission probability  $T(E)$ through the barrier causes the $\sigma_+$ spins to have a higher transmission probability through the QPC (Fig.~\ref{fig:spin_accumulation}e) than $\sigma_-$ even at zero field. In both the linear and non-linear regimes, the charge current through the QPC (Figs.~\ref{fig:spin_accumulation}f and g) causes a restoring voltage $V_3$ to maintain zero net charge current through the QPC with terminal 3 set as a floating probe. While the drive current $I_{\mathrm{sd}}$ oscillates at a frequency $\omega$, the linear and non-linear signals oscillate at the first and second harmonics of $V_3$, i.e. $V_3(\omega)$ and  $V_3(2\omega)$ respectively.
		
\textit{Theoretical analysis.}  Using the transmission probability $T(E) \equiv T(E,B)$ for a QPC \cite{Buettiker1990} (see also Sec.~S1 of the Supplementary Material \cite{Note1}), in the linear regime, the spin signal is proportional to the Zeeman splitting of the one-dimensional subbands. This gives rise to a three-terminal voltage $V_3(\omega)\equiv V_3(\omega,B)$ asymmetric in $B$. The asymmetry $\partial_{B} V_3(\omega)|_{B=0}$ is \cite{Stano2012}:
\begin{equation}
	\partial_{B} V_3(\omega)|_{B=0} = - \frac{\sigma g \mu_B}{2}  \left[\frac{2e}{h} \int dE \left(-\partial_E f(E) \right) \partial_E T(E)\right] \delta \mu_s,
	\label{eq:linear_s2c_general}
\end{equation}
where $\sigma$ is the sign of the spin accumulation, $g$ is the in-plane $g-$factor, $\mu_B$ is the Bohr magneton, and $f(E)$ is the Fermi-Dirac distribution. Eq.~\eqref{eq:linear_s2c_general} allows one to quantify the spin accumulation from the voltage asymmetry. The spin current through the QPC is \cite{Stano2011,Nichele2015}
\begin{equation}
	I_{\mathrm{spin,linear}} \simeq  \frac{2\hbar \Omega_{\mathrm{qpc}}}{\pi g \mu_B} \frac{e^2}{h} \partial_{B} V_3(\omega) |_{B = 0}
	\label{eq:spin_current_from_V3},
\end{equation}
where $\hbar\Omega_{\mathrm{qpc}}$ is the QPC saddle potential curvature \cite{Buettiker1990}. 

\begin{figure*}[t]
	\begin{center}
		\includegraphics[scale = 0.95,trim={0.6cm 0.3cm 0.3cm 0.3cm}]{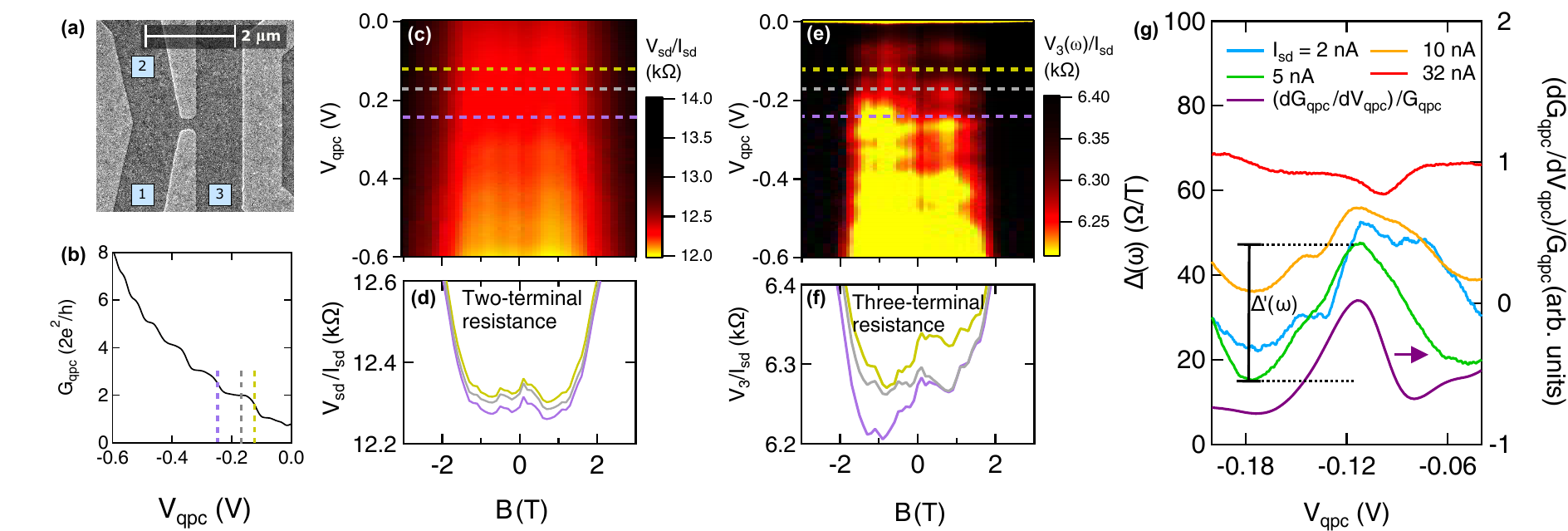}
		\caption{Device image and linear spin-to-charge conversion. (a) A scanning electron microscope image of the device. Light gray regions denote the surface gates, dark gray regions represent the AlGaAs/GaAs heterostructure, while light blue squares depict terminals 1-3. (b) QPC conductance versus QPC gate voltage  $V_{\mathrm{qpc}}$. The dashed lines denote the second and third conductance risers as well as the second conductance plateau.  (c) Color map of the two-terminal resistance $V_{\mathrm{sd}}/I_{\mathrm{sd}}$ across the channel as a function of $V_{\mathrm{qpc}}$ and $B$, showing a symmetric dependence on $B$. (d) Line cuts of $V_{\mathrm{sd}}/I_{\mathrm{sd}}$ from (c) along the second and third QPC conductance risers, as well as along the middle of the second conductance plateau.  (e) Color map of the three-terminal resistance $V_{3}(\omega)/I_{\mathrm{sd}}$ as a function of $V_{\mathrm{qpc}}$ and $B$. Its asymmetry in $B$ indicates a spin accumulation. (f) Line cuts of $V_{3}(\omega)/I_{\mathrm{sd}}$ from (e) along the second and third QPC conductance risers, as well as along the middle of the second conductance plateau. The asymmetry of $V_{3}(\omega)/I_{\mathrm{sd}}$ in $B$ is present at a QPC conductance riser but absent at a plateau. (g) The asymmetry $\Delta(\omega) \equiv (\partial V_3/\partial{B})|_{B=0}/I_{\mathrm{sd}}$ of $V_{3}(\omega)/I_{\mathrm{sd}}$ as a function of $V_{\mathrm{qpc}}$ around the second riser at different $I_{\mathrm{sd}}$. The asymmetry of $V_{3}(\omega)/I_{\mathrm{sd}}$ persists up to $I_{\mathrm{sd}} = 10$ nA, but becomes hard to correlate to the transconductance at 32 nA. The $\Delta(\omega)$ traces are offset by 25 $\Omega/T$ for clarity. The quantity $\Delta'(\omega)$ measures the amplitude of the asymmetry $\Delta(\omega)$ relative to the background.}
		\label{fig:linear_regime_spin_currents_old}
	\end{center}
\end{figure*}

In the non-linear regime, the difference in the transmission probability across the QPC is proportional to $\delta \mu_{s}$. Thus, the non-linear component of the spin signal $V_3$ is quadratic in $\delta\mu_{s}$:
\begin{equation}
V_{3}(2 \omega) = \frac{1}{2} \left[\frac{e}{h} \int dE \left(-\partial_E f(E) \right) \partial_E T(E)\right] (\delta \mu_s) ^ 2.
\label{eq:nonlinear_s2c_general}
\end{equation}
Given that $V_3(2\omega)$ is independent of the sign of $\delta \mu_s$, it is also symmetric in $B$.


Besides quantifying the spin current and accumulation, Eqs.~\eqref{eq:linear_s2c_general}-\eqref{eq:nonlinear_s2c_general} allow us to verify the spin origin of $V_3(\omega)$ and $V_3(2\omega)$ via their dependence on the QPC gate voltage $V_{\mathrm{qpc}}$, $B$, and $I_{\mathrm{sd}}$. Furthermore, since $\partial_E T(E)$ [Eqs.~\eqref{eq:linear_s2c_general} and \eqref{eq:nonlinear_s2c_general}] correlates with the QPC transconductance $\partial {G_{\mathrm{qpc}}}/\partial {V_{\mathrm{qpc}}}$, we expect maximal (no) spin signals when $\partial {G_{\mathrm{qpc}}}/\partial {V_{\mathrm{qpc}}}$ is maximal (minimal), where $G_{\mathrm{qpc}}$ is the QPC conductance.

\textit{Methods.} An image of the device is shown in Fig.~\ref{fig:linear_regime_spin_currents_old}a. The device is made from an AlGaAs/GaAs heterostructure grown on a (100) GaAs substrate. For the measurements presented here, the two-dimensional hole density is $p = 2\times 10^{11}$ cm$^{-2}$, corresponding to a Fermi wavelength $\lambda_F = 56$ nm, a spin-orbit length  $l_{\mathrm{SO}}=35$ nm, and a mobility $\mu=550,000$ cm$^2$ V$^{-1}$ s$^{-1}$ (see Sec.~S2 of the Supplementary Material \footnote{See the Supplementary Material at [URL] for additional details.}). Surface gates define a conducting region in the shape of a `K', with length 4 $\mu$m  and width 1 $\mu$m, whilst the QPC is 370 nm wide and 210 nm long. When the `K-bar' is defined, the conducting channel  in the region is one-dimensional and the transport is ballistic with a mean free path of 4 $\mu$m (see Sec.~S3 of the Supplementary Material \cite{Note1} for details). All measurements were performed in a dilution fridge using standard lock-in techniques with   $\omega = 7$ Hz. 

We send a current $I_{\mathrm{sd}}$ through the drive channel, and measure the resulting two-terminal $V_{\mathrm{sd}} \equiv V_1 - V_2$ and three-terminal voltages $V_{3}(\omega)$  between terminals 2 and 3 (see also Fig.~\ref{fig:spin_accumulation}). Unless otherwise stated, $I_{\mathrm{sd}}$ is kept at 5 nA. Throughout this work, we concentrate our analysis on the second subband. While the first subband is affected by the ``0.7 feature" \cite{Thomas1996,Danneau2008,Micolich2011,Iqbal2013,Bauer2013}, the spin signal is small for higher subbands ($N_{\mathrm{qpc}} > 3$): The conductance quantization is progressively worse for these subbands, diminishing the spin-to-charge conversion efficiency. Fig.~\ref{fig:linear_regime_spin_currents_old}b shows how the QPC conductance is tuned by the QPC gate voltage. The two outer dashed lines mark the second and third conductance risers, where the spin-to-charge conversion should be most pronounced. The middle dashed line locates the second QPC plateau, where the spin-to-charge conversion  should be suppressed. Fig.~\ref{fig:linear_regime_spin_currents_old}c shows the two-terminal resistance $V_{\mathrm{sd}}/I_{\mathrm{sd}}$  across the drive channel as a function of  $V_{\mathrm{qpc}}$ and $B$. As expected from the Onsager reciprocity relation for electrical current in two-terminal systems, $V_{\mathrm{sd}}/I_{\mathrm{sd}}$ is approximately symmetric in $B$ (the QPC is a small perturbation to the drive channel, see Fig.~S4 of the Supplementary Material \cite{Note1}). Fig.~\ref{fig:linear_regime_spin_currents_old}d shows line cuts of Fig.~\ref{fig:linear_regime_spin_currents_old}c at the second and third QPC conductance risers and at the second QPC conductance plateau, confirming that $V_{\mathrm{sd}}$ is approximately symmetric in $B$ regardless of $V_{\mathrm{qpc}}$.
 
\textit{Linear spin-to-charge conversion.}  We now examine the linear three-terminal voltage $V_3(\omega)$. Fig.~\ref{fig:linear_regime_spin_currents_old}e shows $V_3(\omega)/I_{\mathrm{sd}}$ as a function of $V_{\mathrm{qpc}}$ and $B$, demonstrating that $V_3(\omega)$ is generally asymmetric in $B$. The line cuts of Fig.~\ref{fig:linear_regime_spin_currents_old}e shown in Fig.~\ref{fig:linear_regime_spin_currents_old}f reveal that $V_3(\omega)/I_{\mathrm{sd}}$ is asymmetric in $B$ on the second and third QPC conductance risers, but almost symmetric on the middle of the second conductance plateau. This is a crucial observation for the linear spin-to-charge conversion: The asymmetry of $V_3$ with $B$ is expected only if the spin accumulation is present and the QPC transmission is spin-(Zeeman energy) sensitive. At the QPC conductance plateau, although the spin current is still flowing through the QPC, it is not converted to a charge voltage. The asymmetry in $V_3(\omega)$ as a function of $B$ cannot be due to a Hall voltage as the sample was oriented to within $\pm 0.01^{\circ}$ with respect to the magnetic field \cite{Yeoh2010}, so that the out-of-plane magnetic field is always $<0.5$ mT. 

We next quantify the spin accumulation, spin current and the spin-to-charge conversion efficiency. Fig.~\ref{fig:linear_regime_spin_currents_old}g shows the asymmetry $\Delta(\omega) \equiv \frac{1}{I_{\mathrm{sd}}}\frac{\partial V_3(\omega)}{\partial B}\Bigr|_{B=0} \propto I_{\mathrm{spin,linear}} $ of the three-terminal resistance at $I_{\mathrm{sd}} = 2,~5,~10,~32$ nA as a function of $V_{\mathrm{qpc}}$. The asymmetry $\Delta(\omega)$ is obtained by performing a linear fit of $V_3(\omega)$ against $B$ between $-1~ \mathrm{T} \leq B \leq 1$ T in Fig.~\ref{fig:linear_regime_spin_currents_old}g \footnote{The fitting ranges $|B| \leq 0.5, 0.75, 1, 1.25$ T give the same peak positions (see Sec.~S5 of the Supplementary Material \cite{Note2}).}. There is a clear correlation between $\Delta(\omega)$ and $\partial_{V_{\mathrm{qpc}}} G_{\mathrm{qpc}}/G_{\mathrm{qpc}}$, which indicates linear spin-to-charge conversion (Eq.~\ref{eq:linear_s2c_general}) for currents up to $I_{\mathrm{sd}}=10$ nA. The spin signal is suppressed at large $I_{\mathrm{sd}}$ (e.g. at  $I_{\mathrm{sd}} = 32$ nA), possibly due to averaging out of spin accumulations at different energies \cite{Nichele2015}. 

Using the results in Fig.~\ref{fig:linear_regime_spin_currents_old}g and experimental parameters $I_{\mathrm{sd}} = 5$ nA, $g = 0.38 \pm 0.01 $, $\hbar \Omega_{\mathrm{qpc}} = (0.17 \pm 0.01)$ meV (see Sec.~S3 of the Supplementary Material), $\Delta(\omega) = 40$ $\mathrm{\Omega}$/T, $N_{\mathrm{drive}} = 14$ (see Sec.~S4 of the Supplementary Material \cite{Note2}) and $N_{\mathrm{qpc}} = 1.5$, we find that the spin accumulation is  $\delta \mu_s = 1~\mu$eV (Eq.~S7) while the spin current is $I_{\mathrm{spin,linear}} = 37$ pA (Eq.~\ref{eq:spin_current_from_V3}).  The spin Hall angle \cite{Nichele2015}, which measures the spin-to-charge conversion efficiency, is  $\Theta \equiv (I_{\mathrm{spin,linear}}/N_{\mathrm{qpc}})/(I_{\mathrm{sd}}/N_{\mathrm{drive}}) = 6.8\%$. While our spin Hall angle falls within the range of previously reported values \cite{Jungwirth2012,Balakrishnan2014a,Nichele2015,Sinova2015}, caution must be exercised in the comparison since $\Theta$ is not only determined by the material but also the device details.

\begin{figure}[t]
	\begin{center}
		\includegraphics[scale = 1.1,trim={0cm 0.3cm 0cm 0cm}]{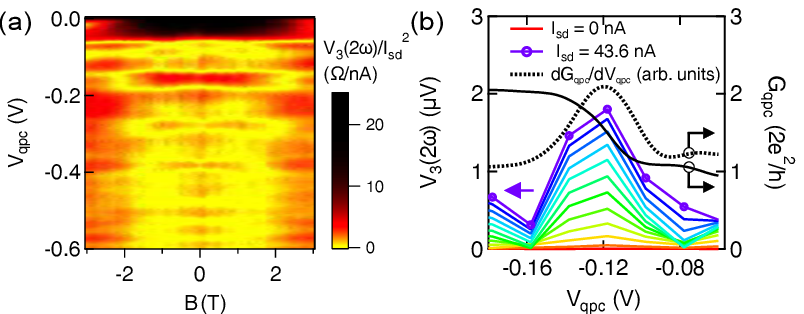}
		\caption{(a) The non-linear resistance $V_3(2\omega)/I_{\mathrm{sd}}^2$ as a function of $V_{\mathrm{qpc}}$ and $B$. (b) The non-linear voltage $V_3(2\omega)$ as a function of $V_{\mathrm{qpc}}$ for various $I_{\mathrm{sd}}$  is proportional to the QPC transconductance at $B = 0$ T.}
		\label{fig:nonlinear_regime_spin_currents_old}
	\end{center}
\end{figure}

\textit{Non-linear spin-to-charge conversion.} Now that we have established evidence for spin-to-charge conversion in the linear regime, we show that it also occurs in the non-linear regime. As before, we evaluate the dependence of the non-linear signal $V_{3}(2\omega)$ on $B$, $V_{\mathrm{qpc}}$, and $I_{\mathrm{sd}}$. Fig.~\ref{fig:nonlinear_regime_spin_currents_old}a shows a color map of the non-linear resistance $V_{3}(2\omega)/I_{\mathrm{sd}}^2$ as a function of $B$ and $V_{\mathrm{qpc}}$. The non-linear signal  $V_3(2\omega)$ is symmetric in $B$, contrasting with the linear signal $V_3(\omega)$ (Fig.~\ref{fig:linear_regime_spin_currents_old}e), and in line with Eq.~\eqref{eq:nonlinear_s2c_general}. Next, we examine the dependence of  $V_{\mathrm{qpc}}$ at $0 \leq I_{\mathrm{sd}} \leq 44.1$ nA at $B = 0$ T (Fig.~\ref{fig:nonlinear_regime_spin_currents_old}b). The peak in the non-linear signal coincides with the QPC transconductance since $\partial_E T(E)$ is maximal at $T(E) = 1/2$ when $B = 0$ T, consistent with Eq.~\eqref{eq:nonlinear_s2c_general}.

\begin{figure}
	\begin{center}
		\includegraphics[scale = 0.6,trim={0cm 0.5cm 0cm 0cm}]{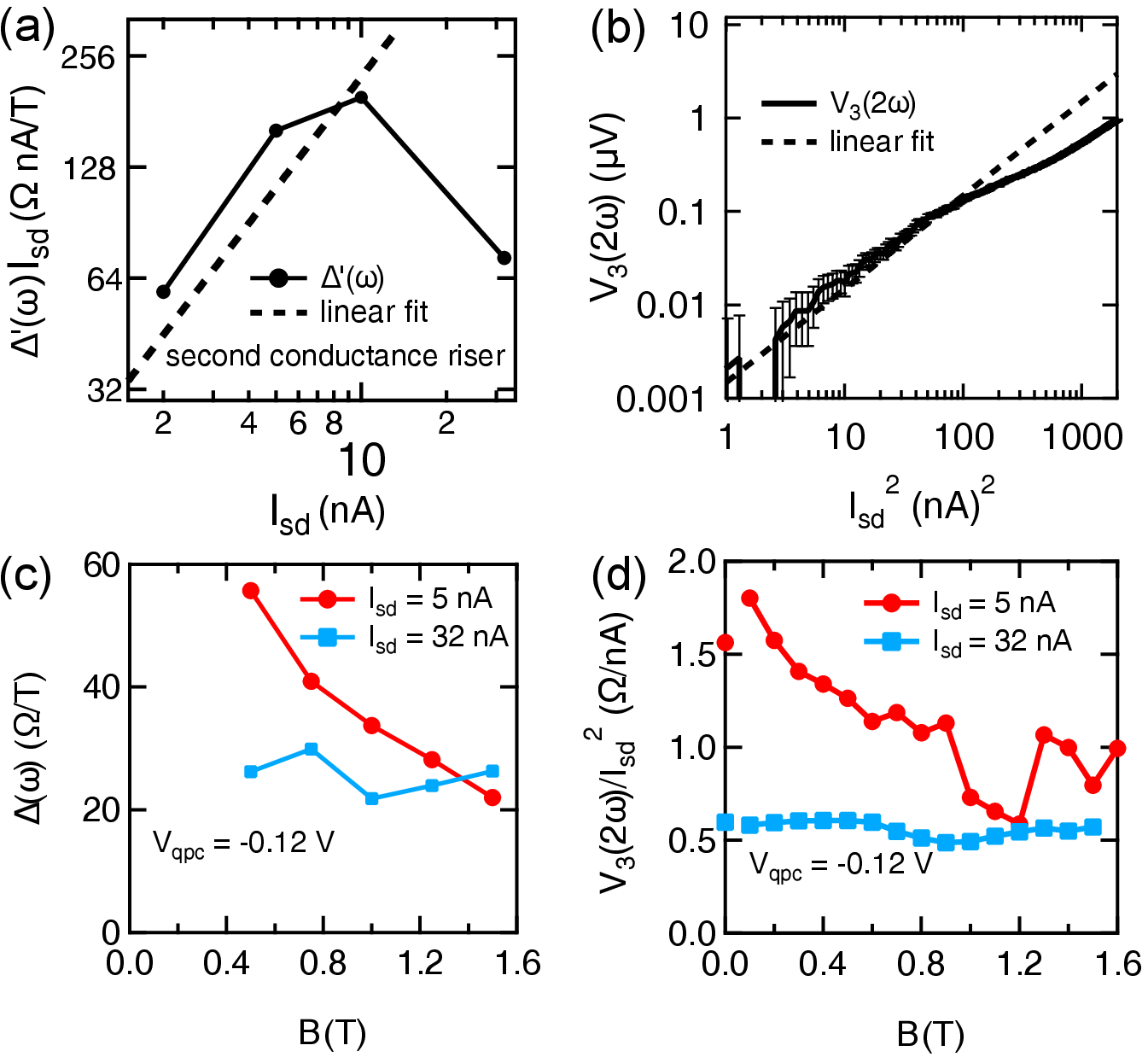}
		\caption{(a) The amplitude $\Delta'(\omega)I_{\mathrm{sd}}$ of the linear signal relative to the background, which is obtained by subtracting the lowest adjacent $\Delta(\omega)$ minimum (see also Fig.~\ref{fig:linear_regime_spin_currents_old}g) from $\Delta(\omega)$ at the second QPC riser. (b) $V_3(2\omega)$ versus $I_{\mathrm{sd}}$ at $B = 0$. The dashed lines in (a) and (b) are guides to the eye, suggesting that the linear and non-linear spin signals saturate at $\sim$5-10 nA. (c) The asymmetry $\Delta(\omega)$ of the linear signal and (d) the non-linear resistance $V_{3}(2\omega)/I_{\mathrm{sd}}^2$ versus $B$. (c) and (d) show that at low excitation currents (e.g. $I_{\mathrm{sd}} = 5$ nA), the spin signal gradually decreases as a function of $B$, whereas at high excitation currents (e.g. $I_{\mathrm{sd}} = 32$ nA), it is almost unaffected by $B$. The difference in the $B-$dependence of the low and high $I_{\mathrm{sd}}$ signals suggests that at low $I_{\mathrm{sd}}$ the three-terminal voltages are of spin origin.}
		\label{fig:current_and_Bpar_dependence}
	\end{center}
\end{figure}

We next compare the linear and non-linear signals. Fig.~\ref{fig:current_and_Bpar_dependence}a shows the amplitude $\Delta'(\omega)$ of the linear signal relative to the background, i.e. the value of $\Delta(\omega)$ at the second subband subtracted by the lowest minimum (see Fig.~\ref{fig:linear_regime_spin_currents_old}g), against $I_{\mathrm{sd}}$. The spin current is linear in $I_{\mathrm{sd}}$ (and hence $\delta \mu_s$) at low excitation currents  ($I_{\mathrm{sd}} \lesssim 5$ nA, see Fig.~\ref{fig:current_and_Bpar_dependence}a). For comparison, Fig.~\ref{fig:current_and_Bpar_dependence}b shows how $V_3(2\omega)$ varies with $I_{\mathrm{sd}}$. We find that the non-linear voltage is proportional to $I_{\mathrm{sd}}^2$ for $I_{\mathrm{sd}} \lesssim 7$ nA. While there is a possibility that Joule heating, which causes thermopower \cite{Appleyard1998,Bakker2010}, could contribute to the second-harmonic response, the fact that both the linear and non-linear signals saturate at similar $I_{\mathrm{sd}}$ ($\approx 5$ nA and $\approx 7$ nA for the linear and non-linear signals, respectively) suggests that they are of a spin origin. 

To further verify the spin origin of the signals, we consider their dependence on $B$ at low ($I_{\mathrm{sd}} = 5$ nA) and high ($I_{\mathrm{sd}} = 32$ nA) excitation currents. At low $I_{\mathrm{sd}}$ (Figs.~\ref{fig:current_and_Bpar_dependence}a and b), both the linear (Fig.~\ref{fig:current_and_Bpar_dependence}c, see also Sec.~S5 of the Supplementary Material \cite{Note2}) and non-linear signals (Fig.~\ref{fig:current_and_Bpar_dependence}d) gradually decrease at $B \gtrsim 1.4$ T, suggesting that a strong magnetic field suppresses the spin accumulation. In contrast, for high $I_{\mathrm{sd}}$, where the spin-to-charge conversion is inefficient \cite{Stano2012}, both the linear and non-linear signals are almost unaffected by the in-plane magnetic field (see also Sec.~S6 of the Supplementary Material \cite{Note2}). The consistency between the linear and non-linear signals confirms the reliability of non-linear spin-to-charge conversion. As non-linear spin-to-charge conversion requires no magnetic field (Figs.~\ref{fig:nonlinear_regime_spin_currents_old}b and \ref{fig:current_and_Bpar_dependence}b), it allows a much faster detection of spin accumulation than linear spin-to-charge conversion \footnote{As an example, the linear spin-to-charge measurements in Fig.~\ref{fig:linear_regime_spin_currents_old} took $\sim30$ hours, as they needed data at both $+B$ and $-B$. The equivalent 	non-linear spin-to-charge conversion, which only involved measuring $V_3$ as a function of $V_{\mathrm{qpc}}$ at $B=0$, took only $\sim 10$ minutes. }.

\textit{Conclusions and outlook.} Using ballistic mesoscopic GaAs holes as a model system, we demonstrate a new all-electrical non-linear technique for spin-to-charge conversion that does not require a magnetic field. We confirm the spin origin of the non-linear signals by calibrating them against linear spin-to-charge conversion. The non-linear spin detection technique allows much faster measurements than linear detection schemes, limited only by the bandwidth of the measurement circuit. Finally, we note that non-linear spin-to-charge conversion is very general: it only requires a spin accumulation regardless of its orientation and an adjacent energy-selective barrier. Our methods should be applicable in materials with very strong spin-orbit interaction such as GaSb, InAs, transition metal dichalcogenides, and topological materials, while its rapid speed will enable time resolved measurements of spin orientation to a 1 ns resolution using radio-frequency techniques.

\textit{Acknowledgment.} The authors would like to thank Heiner Linke and I-Ju Chen for many enlightening discussions. This work was supported by the Australian Research Council under the Discovery Projects scheme and CE170100039. The device 27F8Q was fabricated using the facilities of the New South Wales 279 Node of the Australian National Fabrication Facility. F. Nichele acknowledges support from European Research Commission, grant number 804273.


\begin{thebibliography}{47}%
	\makeatletter
	\providecommand \@ifxundefined [1]{%
		\@ifx{#1\undefined}
	}%
	\providecommand \@ifnum [1]{%
		\ifnum #1\expandafter \@firstoftwo
		\else \expandafter \@secondoftwo
		\fi
	}%
	\providecommand \@ifx [1]{%
		\ifx #1\expandafter \@firstoftwo
		\else \expandafter \@secondoftwo
		\fi
	}%
	\providecommand \natexlab [1]{#1}%
	\providecommand \enquote  [1]{``#1''}%
	\providecommand \bibnamefont  [1]{#1}%
	\providecommand \bibfnamefont [1]{#1}%
	\providecommand \citenamefont [1]{#1}%
	\providecommand \href@noop [0]{\@secondoftwo}%
	\providecommand \href [0]{\begingroup \@sanitize@url \@href}%
	\providecommand \@href[1]{\@@startlink{#1}\@@href}%
	\providecommand \@@href[1]{\endgroup#1\@@endlink}%
	\providecommand \@sanitize@url [0]{\catcode `\\12\catcode `\$12\catcode
		`\&12\catcode `\#12\catcode `\^12\catcode `\_12\catcode `\%12\relax}%
	\providecommand \@@startlink[1]{}%
	\providecommand \@@endlink[0]{}%
	\providecommand \url  [0]{\begingroup\@sanitize@url \@url }%
	\providecommand \@url [1]{\endgroup\@href {#1}{\urlprefix }}%
	\providecommand \urlprefix  [0]{URL }%
	\providecommand \Eprint [0]{\href }%
	\providecommand \doibase [0]{http://dx.doi.org/}%
	\providecommand \selectlanguage [0]{\@gobble}%
	\providecommand \bibinfo  [0]{\@secondoftwo}%
	\providecommand \bibfield  [0]{\@secondoftwo}%
	\providecommand \translation [1]{[#1]}%
	\providecommand \BibitemOpen [0]{}%
	\providecommand \bibitemStop [0]{}%
	\providecommand \bibitemNoStop [0]{.\EOS\space}%
	\providecommand \EOS [0]{\spacefactor3000\relax}%
	\providecommand \BibitemShut  [1]{\csname bibitem#1\endcsname}%
	\let\auto@bib@innerbib\@empty
	\bibitem [{\citenamefont {Wolf}\ \emph {et~al.}(2001)\citenamefont {Wolf},
		\citenamefont {Awschalom}, \citenamefont {Buhrman}, \citenamefont {Daughton},
		\citenamefont {von Moln{\'{a}}r}, \citenamefont {Roukes}, \citenamefont
		{Chtchelkanova},\ and\ \citenamefont {Treger}}]{Wolf2001}%
	\BibitemOpen
	\bibfield  {author} {\bibinfo {author} {\bibfnamefont {S.~A.}\ \bibnamefont
			{Wolf}}, \bibinfo {author} {\bibfnamefont {D.~D.}\ \bibnamefont {Awschalom}},
		\bibinfo {author} {\bibfnamefont {R.~A.}\ \bibnamefont {Buhrman}}, \bibinfo
		{author} {\bibfnamefont {J.~M.}\ \bibnamefont {Daughton}}, \bibinfo {author}
		{\bibfnamefont {S.}~\bibnamefont {von Moln{\'{a}}r}}, \bibinfo {author}
		{\bibfnamefont {M.~L.}\ \bibnamefont {Roukes}}, \bibinfo {author}
		{\bibfnamefont {A.~Y.}\ \bibnamefont {Chtchelkanova}}, \ and\ \bibinfo
		{author} {\bibfnamefont {D.~M.}\ \bibnamefont {Treger}},\ }\href@noop {}
	{\bibfield  {journal} {\bibinfo  {journal} {Science}\ }\textbf {\bibinfo
			{volume} {294}},\ \bibinfo {pages} {1488} (\bibinfo {year}
		{2001})}\BibitemShut {NoStop}%
	\bibitem [{\citenamefont {{\v{Z}}uti{\'{c}}}\ \emph {et~al.}(2004)\citenamefont
		{{\v{Z}}uti{\'{c}}}, \citenamefont {Fabian},\ and\ \citenamefont {{Das
				Sarma}}}]{Zutic2004}%
	\BibitemOpen
	\bibfield  {author} {\bibinfo {author} {\bibfnamefont {I.}~\bibnamefont
			{{\v{Z}}uti{\'{c}}}}, \bibinfo {author} {\bibfnamefont {J.}~\bibnamefont
			{Fabian}}, \ and\ \bibinfo {author} {\bibfnamefont {S.}~\bibnamefont {{Das
					Sarma}}},\ }\href@noop {} {\bibfield  {journal} {\bibinfo  {journal} {Rev.
				Mod. Phys.}\ }\textbf {\bibinfo {volume} {76}},\ \bibinfo {pages} {323}
		(\bibinfo {year} {2004})}\BibitemShut {NoStop}%
	\bibitem [{\citenamefont {Han}\ \emph {et~al.}(2018)\citenamefont {Han},
		\citenamefont {Otani},\ and\ \citenamefont {Maekawa}}]{Han2018}%
	\BibitemOpen
	\bibfield  {author} {\bibinfo {author} {\bibfnamefont {W.}~\bibnamefont
			{Han}}, \bibinfo {author} {\bibfnamefont {Y.}~\bibnamefont {Otani}}, \ and\
		\bibinfo {author} {\bibfnamefont {S.}~\bibnamefont {Maekawa}},\ }\href@noop
	{} {\bibfield  {journal} {\bibinfo  {journal} {npj quantum mater.}\ }\textbf
		{\bibinfo {volume} {3}},\ \bibinfo {pages} {1} (\bibinfo {year}
		{2018})}\BibitemShut {NoStop}%
	\bibitem [{\citenamefont {Stern}\ \emph {et~al.}(2006)\citenamefont {Stern},
		\citenamefont {Ghosh}, \citenamefont {Xiang}, \citenamefont {Zhu},
		\citenamefont {Samarth},\ and\ \citenamefont {Awschalom}}]{Stern2006}%
	\BibitemOpen
	\bibfield  {author} {\bibinfo {author} {\bibfnamefont {N.~P.}\ \bibnamefont
			{Stern}}, \bibinfo {author} {\bibfnamefont {S.}~\bibnamefont {Ghosh}},
		\bibinfo {author} {\bibfnamefont {G.}~\bibnamefont {Xiang}}, \bibinfo
		{author} {\bibfnamefont {M.}~\bibnamefont {Zhu}}, \bibinfo {author}
		{\bibfnamefont {N.}~\bibnamefont {Samarth}}, \ and\ \bibinfo {author}
		{\bibfnamefont {D.~D.}\ \bibnamefont {Awschalom}},\ }\href@noop {} {\bibfield
		{journal} {\bibinfo  {journal} {Phys. Rev. Lett.}\ }\textbf {\bibinfo
			{volume} {97}},\ \bibinfo {pages} {126603} (\bibinfo {year}
		{2006})}\BibitemShut {NoStop}%
	\bibitem [{\citenamefont {Chang}\ \emph {et~al.}(2007)\citenamefont {Chang},
		\citenamefont {Chen}, \citenamefont {Chen}, \citenamefont {Hong},
		\citenamefont {Tsai}, \citenamefont {Chen},\ and\ \citenamefont
		{Guo}}]{Chang2007}%
	\BibitemOpen
	\bibfield  {author} {\bibinfo {author} {\bibfnamefont {H.~J.}\ \bibnamefont
			{Chang}}, \bibinfo {author} {\bibfnamefont {T.~W.}\ \bibnamefont {Chen}},
		\bibinfo {author} {\bibfnamefont {J.~W.}\ \bibnamefont {Chen}}, \bibinfo
		{author} {\bibfnamefont {W.~C.}\ \bibnamefont {Hong}}, \bibinfo {author}
		{\bibfnamefont {W.~C.}\ \bibnamefont {Tsai}}, \bibinfo {author}
		{\bibfnamefont {Y.~F.}\ \bibnamefont {Chen}}, \ and\ \bibinfo {author}
		{\bibfnamefont {G.~Y.}\ \bibnamefont {Guo}},\ }\href@noop {} {\bibfield
		{journal} {\bibinfo  {journal} {Phys. Rev. Lett.}\ }\textbf {\bibinfo
			{volume} {98}},\ \bibinfo {pages} {136403} (\bibinfo {year}
		{2007})}\BibitemShut {NoStop}%
	\bibitem [{\citenamefont {Matsuzaka}\ \emph {et~al.}(2009)\citenamefont
		{Matsuzaka}, \citenamefont {Ohno},\ and\ \citenamefont
		{Ohno}}]{Matsuzaka2009}%
	\BibitemOpen
	\bibfield  {author} {\bibinfo {author} {\bibfnamefont {S.}~\bibnamefont
			{Matsuzaka}}, \bibinfo {author} {\bibfnamefont {Y.}~\bibnamefont {Ohno}}, \
		and\ \bibinfo {author} {\bibfnamefont {H.}~\bibnamefont {Ohno}},\ }\href@noop
	{} {\bibfield  {journal} {\bibinfo  {journal} {Phys. Rev. B}\ }\textbf
		{\bibinfo {volume} {80}},\ \bibinfo {pages} {241305(R)} (\bibinfo {year}
		{2009})}\BibitemShut {NoStop}%
	\bibitem [{\citenamefont {Ando}\ \emph {et~al.}(2010)\citenamefont {Ando},
		\citenamefont {Morikawa}, \citenamefont {Trypiniotis}, \citenamefont
		{Fujikawa}, \citenamefont {Barnes},\ and\ \citenamefont {Saitoh}}]{Ando2010}%
	\BibitemOpen
	\bibfield  {author} {\bibinfo {author} {\bibfnamefont {K.}~\bibnamefont
			{Ando}}, \bibinfo {author} {\bibfnamefont {M.}~\bibnamefont {Morikawa}},
		\bibinfo {author} {\bibfnamefont {T.}~\bibnamefont {Trypiniotis}}, \bibinfo
		{author} {\bibfnamefont {Y.}~\bibnamefont {Fujikawa}}, \bibinfo {author}
		{\bibfnamefont {C.~H.~W.}\ \bibnamefont {Barnes}}, \ and\ \bibinfo {author}
		{\bibfnamefont {E.}~\bibnamefont {Saitoh}},\ }\href {\doibase
		10.1063/1.3418441} {\bibfield  {journal} {\bibinfo  {journal} {Appl. Phys.
				Lett.}\ }\textbf {\bibinfo {volume} {86}},\ \bibinfo {pages} {082502}
		(\bibinfo {year} {2010})}\BibitemShut {NoStop}%
	\bibitem [{\citenamefont {Wunderlich}\ \emph {et~al.}(2010)\citenamefont
		{Wunderlich}, \citenamefont {Park}, \citenamefont {Irvine}, \citenamefont
		{Z{\^{a}}rbo}, \citenamefont {Rozkotov{\'{a}}}, \citenamefont {Nemec},
		\citenamefont {Nov{\'{a}}k}, \citenamefont {Sinova},\ and\ \citenamefont
		{Jungwirth}}]{Wunderlich2010}%
	\BibitemOpen
	\bibfield  {author} {\bibinfo {author} {\bibfnamefont {J.}~\bibnamefont
			{Wunderlich}}, \bibinfo {author} {\bibfnamefont {B.-G.}\ \bibnamefont
			{Park}}, \bibinfo {author} {\bibfnamefont {A.~C.}\ \bibnamefont {Irvine}},
		\bibinfo {author} {\bibfnamefont {L.~P.}\ \bibnamefont {Z{\^{a}}rbo}},
		\bibinfo {author} {\bibfnamefont {E.}~\bibnamefont {Rozkotov{\'{a}}}},
		\bibinfo {author} {\bibfnamefont {P.}~\bibnamefont {Nemec}}, \bibinfo
		{author} {\bibfnamefont {V.}~\bibnamefont {Nov{\'{a}}k}}, \bibinfo {author}
		{\bibfnamefont {J.}~\bibnamefont {Sinova}}, \ and\ \bibinfo {author}
		{\bibfnamefont {T.}~\bibnamefont {Jungwirth}},\ }\href {\doibase
		10.1126/science.1195816} {\bibfield  {journal} {\bibinfo  {journal}
			{Science}\ }\textbf {\bibinfo {volume} {330}},\ \bibinfo {pages} {1801}
		(\bibinfo {year} {2010})}\BibitemShut {NoStop}%
	\bibitem [{\citenamefont {Br{\"{u}}ne}\ \emph {et~al.}(2010)\citenamefont
		{Br{\"{u}}ne}, \citenamefont {Roth}, \citenamefont {Novik}, \citenamefont
		{K{\"{o}}nig}, \citenamefont {Buhmann}, \citenamefont {Hankiewicz},
		\citenamefont {Hanke},\ and\ \citenamefont {Sinova}}]{Bruene2010}%
	\BibitemOpen
	\bibfield  {author} {\bibinfo {author} {\bibfnamefont {C.}~\bibnamefont
			{Br{\"{u}}ne}}, \bibinfo {author} {\bibfnamefont {A.}~\bibnamefont {Roth}},
		\bibinfo {author} {\bibfnamefont {E.~G.}\ \bibnamefont {Novik}}, \bibinfo
		{author} {\bibfnamefont {M.}~\bibnamefont {K{\"{o}}nig}}, \bibinfo {author}
		{\bibfnamefont {H.}~\bibnamefont {Buhmann}}, \bibinfo {author} {\bibfnamefont
			{E.~M.}\ \bibnamefont {Hankiewicz}}, \bibinfo {author} {\bibfnamefont
			{W.}~\bibnamefont {Hanke}}, \ and\ \bibinfo {author} {\bibfnamefont
			{J.}~\bibnamefont {Sinova}},\ }\href {\doibase 10.1038/nphys1655} {\bibfield
		{journal} {\bibinfo  {journal} {Nat. Phys.}\ }\textbf {\bibinfo {volume}
			{6}},\ \bibinfo {pages} {448} (\bibinfo {year} {2010})}\BibitemShut {NoStop}%
	\bibitem [{\citenamefont {Balakrishnan}\ \emph {et~al.}(2013)\citenamefont
		{Balakrishnan}, \citenamefont {{Kok Wai Koon}}, \citenamefont {Jaiswal},
		\citenamefont {{Castro Neto}},\ and\ \citenamefont
		{{\"{O}}zyilmaz}}]{Balakrishnan2013a}%
	\BibitemOpen
	\bibfield  {author} {\bibinfo {author} {\bibfnamefont {J.}~\bibnamefont
			{Balakrishnan}}, \bibinfo {author} {\bibfnamefont {G.}~\bibnamefont {{Kok Wai
					Koon}}}, \bibinfo {author} {\bibfnamefont {M.}~\bibnamefont {Jaiswal}},
		\bibinfo {author} {\bibfnamefont {A.~H.}\ \bibnamefont {{Castro Neto}}}, \
		and\ \bibinfo {author} {\bibfnamefont {B.}~\bibnamefont {{\"{O}}zyilmaz}},\
	}\href {\doibase 10.1038/nphys2576} {\bibfield  {journal} {\bibinfo
			{journal} {Nat. Phys.}\ }\textbf {\bibinfo {volume} {9}},\ \bibinfo {pages}
		{284} (\bibinfo {year} {2013})}\BibitemShut {NoStop}%
	\bibitem [{\citenamefont {Choi}\ \emph {et~al.}(2015)\citenamefont {Choi},
		\citenamefont {Kim}, \citenamefont {Chang}, \citenamefont {Han},
		\citenamefont {Koo},\ and\ \citenamefont {Johnson}}]{Choi2015}%
	\BibitemOpen
	\bibfield  {author} {\bibinfo {author} {\bibfnamefont {W.~Y.}\ \bibnamefont
			{Choi}}, \bibinfo {author} {\bibfnamefont {H.-J.}\ \bibnamefont {Kim}},
		\bibinfo {author} {\bibfnamefont {J.}~\bibnamefont {Chang}}, \bibinfo
		{author} {\bibfnamefont {S.~H.}\ \bibnamefont {Han}}, \bibinfo {author}
		{\bibfnamefont {H.~C.}\ \bibnamefont {Koo}}, \ and\ \bibinfo {author}
		{\bibfnamefont {M.}~\bibnamefont {Johnson}},\ }\href@noop {} {\bibfield
		{journal} {\bibinfo  {journal} {Nature Nanotechnol.}\ }\textbf {\bibinfo
			{volume} {10}},\ \bibinfo {pages} {666} (\bibinfo {year} {2015})}\BibitemShut
	{NoStop}%
	\bibitem [{\citenamefont {Bottegoni}\ \emph {et~al.}(2017)\citenamefont
		{Bottegoni}, \citenamefont {Zucchetti}, \citenamefont {Dal~Conte},
		\citenamefont {Frigerio}, \citenamefont {Carpene}, \citenamefont {Vergnaud},
		\citenamefont {Jamet}, \citenamefont {Isella}, \citenamefont {Ciccacci},
		\citenamefont {Cerullo},\ and\ \citenamefont {Finazzi}}]{Bottegoni2017}%
	\BibitemOpen
	\bibfield  {author} {\bibinfo {author} {\bibfnamefont {F.}~\bibnamefont
			{Bottegoni}}, \bibinfo {author} {\bibfnamefont {C.}~\bibnamefont
			{Zucchetti}}, \bibinfo {author} {\bibfnamefont {S.}~\bibnamefont
			{Dal~Conte}}, \bibinfo {author} {\bibfnamefont {J.}~\bibnamefont {Frigerio}},
		\bibinfo {author} {\bibfnamefont {E.}~\bibnamefont {Carpene}}, \bibinfo
		{author} {\bibfnamefont {C.}~\bibnamefont {Vergnaud}}, \bibinfo {author}
		{\bibfnamefont {M.}~\bibnamefont {Jamet}}, \bibinfo {author} {\bibfnamefont
			{G.}~\bibnamefont {Isella}}, \bibinfo {author} {\bibfnamefont
			{F.}~\bibnamefont {Ciccacci}}, \bibinfo {author} {\bibfnamefont
			{G.}~\bibnamefont {Cerullo}}, \ and\ \bibinfo {author} {\bibfnamefont
			{M.}~\bibnamefont {Finazzi}},\ }\href@noop {} {\bibfield  {journal} {\bibinfo
			{journal} {Phys. Rev. Lett.}\ }\textbf {\bibinfo {volume} {118}},\ \bibinfo
		{pages} {167402} (\bibinfo {year} {2017})}\BibitemShut {NoStop}%
	\bibitem [{\citenamefont {Gerlach}\ and\ \citenamefont
		{Stern}(1922)}]{SternGerlach1922}%
	\BibitemOpen
	\bibfield  {author} {\bibinfo {author} {\bibfnamefont {W.}~\bibnamefont
			{Gerlach}}\ and\ \bibinfo {author} {\bibfnamefont {O.}~\bibnamefont
			{Stern}},\ }\href@noop {} {\bibfield  {journal} {\bibinfo  {journal} {Z.
				Phys.}\ }\textbf {\bibinfo {volume} {9}},\ \bibinfo {pages} {349} (\bibinfo
		{year} {1922})}\BibitemShut {NoStop}%
	\bibitem [{\citenamefont {Chesi}\ \emph {et~al.}(2011)\citenamefont {Chesi},
		\citenamefont {Giuliani}, \citenamefont {Rokhinson}, \citenamefont
		{Pfeiffer}, \citenamefont {West},\ and\ \citenamefont {W.}}]{Chesi2011}%
	\BibitemOpen
	\bibfield  {author} {\bibinfo {author} {\bibfnamefont {S.}~\bibnamefont
			{Chesi}}, \bibinfo {author} {\bibfnamefont {G.~F.}\ \bibnamefont {Giuliani}},
		\bibinfo {author} {\bibfnamefont {L.~P.}\ \bibnamefont {Rokhinson}}, \bibinfo
		{author} {\bibfnamefont {L.~N.}\ \bibnamefont {Pfeiffer}}, \ and\ \bibinfo {author} {\bibfnamefont
			{K.~W.}}\ \bibnamefont{West} }
		\href@noop {} {\bibfield  {journal} {\bibinfo
			{journal} {Phys. Rev. Lett.}\ }\textbf {\bibinfo {volume} {106}},\ \bibinfo
		{pages} {236601} (\bibinfo {year} {2011})}\BibitemShut {NoStop}%
	\bibitem [{\citenamefont {Nichele}\ \emph {et~al.}(2015)\citenamefont
		{Nichele}, \citenamefont {Hennel}, \citenamefont {Pietsch}, \citenamefont
		{Wegscheider}, \citenamefont {Stano}, \citenamefont {Jacquod}, \citenamefont
		{Ihn},\ and\ \citenamefont {Ensslin}}]{Nichele2015}%
	\BibitemOpen
	\bibfield  {author} {\bibinfo {author} {\bibfnamefont {F.}~\bibnamefont
			{Nichele}}, \bibinfo {author} {\bibfnamefont {S.}~\bibnamefont {Hennel}},
		\bibinfo {author} {\bibfnamefont {P.}~\bibnamefont {Pietsch}}, \bibinfo
		{author} {\bibfnamefont {W.}~\bibnamefont {Wegscheider}}, \bibinfo {author}
		{\bibfnamefont {P.}~\bibnamefont {Stano}}, \bibinfo {author} {\bibfnamefont
			{P.}~\bibnamefont {Jacquod}}, \bibinfo {author} {\bibfnamefont
			{T.}~\bibnamefont {Ihn}}, \ and\ \bibinfo {author} {\bibfnamefont
			{K.}~\bibnamefont {Ensslin}},\ }\href {\doibase
		10.1103/PhysRevLett.114.206601} {\bibfield  {journal} {\bibinfo  {journal}
			{Phys. Rev. Lett.}\ }\textbf {\bibinfo {volume} {114}},\ \bibinfo {pages}
		{206601} (\bibinfo {year} {2015})}\BibitemShut {NoStop}%
	\bibitem [{\citenamefont {Hilton}\ and\ \citenamefont
		{Tang}(2002)}]{Hilton2002}%
	\BibitemOpen
	\bibfield  {author} {\bibinfo {author} {\bibfnamefont {D.~J.}\ \bibnamefont
			{Hilton}}\ and\ \bibinfo {author} {\bibfnamefont {C.~L.}\ \bibnamefont
			{Tang}},\ }\href@noop {} {\bibfield  {journal} {\bibinfo  {journal} {Phys.
				Rev. Lett.}\ }\textbf {\bibinfo {volume} {89}},\ \bibinfo {pages} {146601}
		(\bibinfo {year} {2002})}\BibitemShut {NoStop}%
	\bibitem [{\citenamefont {Akhgar}\ \emph {et~al.}(2016)\citenamefont {Akhgar},
		\citenamefont {Klochan}, \citenamefont {Willems~van Beveren}, \citenamefont
		{Edmonds}, \citenamefont {Maier}, \citenamefont {Spencer}, \citenamefont
		{McCallum}, \citenamefont {Ley}, \citenamefont {Hamilton},\ and\
		\citenamefont {Pakes}}]{Akhgar2016}%
	\BibitemOpen
	\bibfield  {author} {\bibinfo {author} {\bibfnamefont {G.}~\bibnamefont
			{Akhgar}}, \bibinfo {author} {\bibfnamefont {O.}~\bibnamefont {Klochan}},
		\bibinfo {author} {\bibfnamefont {L.~H.}\ \bibnamefont {Willems~van
				Beveren}}, \bibinfo {author} {\bibfnamefont {M.~T.}\ \bibnamefont {Edmonds}},
		\bibinfo {author} {\bibfnamefont {F.}~\bibnamefont {Maier}}, \bibinfo
		{author} {\bibfnamefont {B.~J.}\ \bibnamefont {Spencer}}, \bibinfo {author}
		{\bibfnamefont {J.~C.}\ \bibnamefont {McCallum}}, \bibinfo {author}
		{\bibfnamefont {L.}~\bibnamefont {Ley}}, \bibinfo {author} {\bibfnamefont
			{A.~R.}\ \bibnamefont {Hamilton}}, \ and\ \bibinfo {author} {\bibfnamefont
			{C.~I.}\ \bibnamefont {Pakes}},\ }\href@noop {} {\bibfield  {journal}
		{\bibinfo  {journal} {Nano Letters}\ }\textbf {\bibinfo {volume} {16}},\
		\bibinfo {pages} {3768} (\bibinfo {year} {2016})}\BibitemShut {NoStop}%
	\bibitem [{\citenamefont {Wang}\ \emph {et~al.}(2016)\citenamefont {Wang},
		\citenamefont {Klochan}, \citenamefont {Hung}, \citenamefont {Culcer},
		\citenamefont {Farrer}, \citenamefont {Ritchie},\ and\ \citenamefont
		{Hamilton}}]{Wang2016}%
	\BibitemOpen
	\bibfield  {author} {\bibinfo {author} {\bibfnamefont {D.~Q.}\ \bibnamefont
			{Wang}}, \bibinfo {author} {\bibfnamefont {O.}~\bibnamefont {Klochan}},
		\bibinfo {author} {\bibfnamefont {J.-T.}\ \bibnamefont {Hung}}, \bibinfo
		{author} {\bibfnamefont {D.}~\bibnamefont {Culcer}}, \bibinfo {author}
		{\bibfnamefont {I.}~\bibnamefont {Farrer}}, \bibinfo {author} {\bibfnamefont
			{D.~A.}\ \bibnamefont {Ritchie}}, \ and\ \bibinfo {author} {\bibfnamefont
			{A.~R.}\ \bibnamefont {Hamilton}},\ }\href {\doibase
		10.1021/acs.nanolett.6b03752} {\bibfield  {journal} {\bibinfo  {journal}
			{Nano Lett.}\ }\textbf {\bibinfo {volume} {16}},\ \bibinfo {pages} {7685}
		(\bibinfo {year} {2016})}\BibitemShut {NoStop}%
	\bibitem [{\citenamefont {Srinivasan}\ \emph {et~al.}(2017)\citenamefont
		{Srinivasan}, \citenamefont {Miserev}, \citenamefont {Hudson}, \citenamefont
		{Klochan}, \citenamefont {Muraki}, \citenamefont {Hirayama}, \citenamefont
		{Reuter}, \citenamefont {Wieck}, \citenamefont {Sushkov},\ and\ \citenamefont
		{Hamilton}}]{Srinivasan2017}%
	\BibitemOpen
	\bibfield  {author} {\bibinfo {author} {\bibfnamefont {A.}~\bibnamefont
			{Srinivasan}}, \bibinfo {author} {\bibfnamefont {D.~S.}\ \bibnamefont
			{Miserev}}, \bibinfo {author} {\bibfnamefont {K.~L.}\ \bibnamefont {Hudson}},
		\bibinfo {author} {\bibfnamefont {O.}~\bibnamefont {Klochan}}, \bibinfo
		{author} {\bibfnamefont {K.}~\bibnamefont {Muraki}}, \bibinfo {author}
		{\bibfnamefont {Y.}~\bibnamefont {Hirayama}}, \bibinfo {author}
		{\bibfnamefont {D.}~\bibnamefont {Reuter}}, \bibinfo {author} {\bibfnamefont
			{A.~D.}\ \bibnamefont {Wieck}}, \bibinfo {author} {\bibfnamefont {O.~P.}\
			\bibnamefont {Sushkov}}, \ and\ \bibinfo {author} {\bibfnamefont {A.~R.}\
			\bibnamefont {Hamilton}},\ }\href {\doibase 10.1103/PhysRevLett.118.146801}
	{\bibfield  {journal} {\bibinfo  {journal} {Phys. Rev. Lett.}\ }\textbf
		{\bibinfo {volume} {118}},\ \bibinfo {pages} {146801} (\bibinfo {year}
		{2017})}\BibitemShut {NoStop}%
	\bibitem [{\citenamefont {Marcellina}\ \emph {et~al.}(2018)\citenamefont
		{Marcellina}, \citenamefont {Srinivasan}, \citenamefont {Miserev},
		\citenamefont {Croxall}, \citenamefont {Ritchie}, \citenamefont {Farrer},
		\citenamefont {Sushkov}, \citenamefont {Culcer},\ and\ \citenamefont
		{Hamilton}}]{Marcellina2018}%
	\BibitemOpen
	\bibfield  {author} {\bibinfo {author} {\bibfnamefont {E.}~\bibnamefont
			{Marcellina}}, \bibinfo {author} {\bibfnamefont {A.}~\bibnamefont
			{Srinivasan}}, \bibinfo {author} {\bibfnamefont {D.~S.}\ \bibnamefont
			{Miserev}}, \bibinfo {author} {\bibfnamefont {A.~F.}\ \bibnamefont
			{Croxall}}, \bibinfo {author} {\bibfnamefont {D.~A.}\ \bibnamefont
			{Ritchie}}, \bibinfo {author} {\bibfnamefont {I.}~\bibnamefont {Farrer}},
		\bibinfo {author} {\bibfnamefont {O.~P.}\ \bibnamefont {Sushkov}}, \bibinfo
		{author} {\bibfnamefont {D.}~\bibnamefont {Culcer}}, \ and\ \bibinfo {author}
		{\bibfnamefont {A.~R.}\ \bibnamefont {Hamilton}},\ }\href@noop {} {\bibfield
		{journal} {\bibinfo  {journal} {Phys. Rev. Lett.}\ }\textbf {\bibinfo
			{volume} {121}},\ \bibinfo {pages} {077701} (\bibinfo {year}
		{2018})}\BibitemShut {NoStop}%
	\bibitem [{\citenamefont {Hendrickx}\ \emph {et~al.}(2018)\citenamefont
		{Hendrickx}, \citenamefont {Franke}, \citenamefont {Sammak}, \citenamefont
		{Kouwenhoven}, \citenamefont {Sabbagh}, \citenamefont {Yeoh}, \citenamefont
		{Li}, \citenamefont {Tagliaferri}, \citenamefont {Virgillo}, \citenamefont
		{Capellini}, \citenamefont {Scappucci},\ and\ \citenamefont
		{Veldhorst}}]{Hendrickx2018}%
	\BibitemOpen
	\bibfield  {author} {\bibinfo {author} {\bibfnamefont {N.~W.}\ \bibnamefont
			{Hendrickx}}, \bibinfo {author} {\bibfnamefont {D.~P.}\ \bibnamefont
			{Franke}}, \bibinfo {author} {\bibfnamefont {A.}~\bibnamefont {Sammak}},
		\bibinfo {author} {\bibfnamefont {M.}~\bibnamefont {Kouwenhoven}}, \bibinfo
		{author} {\bibfnamefont {D.}~\bibnamefont {Sabbagh}}, \bibinfo {author}
		{\bibfnamefont {L.}~\bibnamefont {Yeoh}}, \bibinfo {author} {\bibfnamefont
			{R.}~\bibnamefont {Li}}, \bibinfo {author} {\bibfnamefont {M.~L.~V.}\
			\bibnamefont {Tagliaferri}}, \bibinfo {author} {\bibfnamefont
			{M.}~\bibnamefont {Virgillo}}, \bibinfo {author} {\bibfnamefont
			{G.}~\bibnamefont {Capellini}}, \bibinfo {author} {\bibfnamefont
			{G.}~\bibnamefont {Scappucci}}, \ and\ \bibinfo {author} {\bibfnamefont
			{M.}~\bibnamefont {Veldhorst}},\ }\href@noop {} {\bibfield  {journal}
		{\bibinfo  {journal} {Nat. Commun.}\ }\textbf {\bibinfo {volume} {9}},\
		\bibinfo {pages} {1} (\bibinfo {year} {2018})}\BibitemShut {NoStop}%
	\bibitem [{\citenamefont {Watzinger}\ \emph {et~al.}(2018)\citenamefont
		{Watzinger}, \citenamefont {Kuku\v{c}ka}, \citenamefont {Vuku\v{s}i\'{c}},
		\citenamefont {Gao}, \citenamefont {Wang}, \citenamefont {Sc\"{a}ffler},\
		and\ \citenamefont {Katsaros}}]{Watzinger2018}%
	\BibitemOpen
	\bibfield  {author} {\bibinfo {author} {\bibfnamefont {H.}~\bibnamefont
			{Watzinger}}, \bibinfo {author} {\bibfnamefont {J.}~\bibnamefont
			{Kuku\v{c}ka}}, \bibinfo {author} {\bibfnamefont {L.}~\bibnamefont
			{Vuku\v{s}i\'{c}}}, \bibinfo {author} {\bibfnamefont {F.}~\bibnamefont
			{Gao}}, \bibinfo {author} {\bibfnamefont {T.}~\bibnamefont {Wang}}, \bibinfo
		{author} {\bibfnamefont {F.}~\bibnamefont {Sc\"{a}ffler}}, \ and\ \bibinfo
		{author} {\bibfnamefont {G.}~\bibnamefont {Katsaros}},\ }\href@noop {}
	{\bibfield  {journal} {\bibinfo  {journal} {Nat. Commun.}\ }\textbf {\bibinfo
			{volume} {9}},\ \bibinfo {pages} {1} (\bibinfo {year} {2018})}\BibitemShut
	{NoStop}%
	\bibitem [{\citenamefont {Crippa}\ \emph {et~al.}(2018)\citenamefont {Crippa},
		\citenamefont {Maurand}, \citenamefont {Bourdet}, \citenamefont
		{Kotekar-Patil}, \citenamefont {Amisse}, \citenamefont {Jehl}, \citenamefont
		{Sanquer}, \citenamefont {Lavi\'{e}ville}, \citenamefont {Bohuslavskyi},
		\citenamefont {Hutin}, \citenamefont {Barraud}, \citenamefont {Vinet},
		\citenamefont {Niquet},\ and\ \citenamefont {de~Franceschi}}]{Crippa2018}%
	\BibitemOpen
	\bibfield  {author} {\bibinfo {author} {\bibfnamefont {A.}~\bibnamefont
			{Crippa}}, \bibinfo {author} {\bibfnamefont {R.}~\bibnamefont {Maurand}},
		\bibinfo {author} {\bibfnamefont {L.}~\bibnamefont {Bourdet}}, \bibinfo
		{author} {\bibfnamefont {D.}~\bibnamefont {Kotekar-Patil}}, \bibinfo {author}
		{\bibfnamefont {A.}~\bibnamefont {Amisse}}, \bibinfo {author} {\bibfnamefont
			{X.}~\bibnamefont {Jehl}}, \bibinfo {author} {\bibfnamefont {M.}~\bibnamefont
			{Sanquer}}, \bibinfo {author} {\bibfnamefont {R.}~\bibnamefont
			{Lavi\'{e}ville}}, \bibinfo {author} {\bibfnamefont{H.}~\bibnamefont {Bohuslavskyi}}, \bibinfo
		{author} {\bibfnamefont {L.}~\bibnamefont {Hutin}}, \bibinfo {author}
		{\bibfnamefont {S.}~\bibnamefont {Barraud}}, \bibinfo {author} {\bibfnamefont
			{M.}~\bibnamefont {Vinet}}, \bibinfo {author} {\bibfnamefont {Y.-M.}\
			\bibnamefont {Niquet}}, \ and\ \bibinfo {author} {\bibfnamefont
			{S.}~\bibnamefont {De~Franceschi}},\ }\href@noop {} {\bibfield  {journal}
		{\bibinfo  {journal} {Phys. Rev. Lett.}\ }\textbf {\bibinfo {volume} {120}},\
		\bibinfo {pages} {137702} (\bibinfo {year} {2018})}\BibitemShut {NoStop}%
	\bibitem [{\citenamefont {Hendrickx}\ \emph {et~al.}(2020)\citenamefont
		{Hendrickx}, \citenamefont {Franke}, \citenamefont {Sammak}, \citenamefont
		{Scappucci},\ and\ \citenamefont {Veldhorst}}]{Hendrickx2020}%
	\BibitemOpen
	\bibfield  {author} {\bibinfo {author} {\bibfnamefont {N.~W.}\ \bibnamefont
			{Hendrickx}}, \bibinfo {author} {\bibfnamefont {D.~P.}\ \bibnamefont
			{Franke}}, \bibinfo {author} {\bibfnamefont {A.}~\bibnamefont {Sammak}},
		\bibinfo {author} {\bibfnamefont {G.}~\bibnamefont {Scappucci}}, \ and\
		\bibinfo {author} {\bibfnamefont {M.}~\bibnamefont {Veldhorst}},\ }\href@noop
	{} {\bibfield  {journal} {\bibinfo  {journal} {Nature}\ }\textbf {\bibinfo
			{volume} {577}},\ \bibinfo {pages} {487} (\bibinfo {year}
		{2020})}\BibitemShut {NoStop}%
	\bibitem [{\citenamefont {Bardarson}\ \emph {et~al.}(2007)\citenamefont
		{Bardarson}, \citenamefont {Adagideli},\ and\ \citenamefont
		{Jacquod}}]{Bardarson2007}%
	\BibitemOpen
	\bibfield  {author} {\bibinfo {author} {\bibfnamefont {J.~H.}\ \bibnamefont
			{Bardarson}}, \bibinfo {author} {\bibfnamefont {I.}~\bibnamefont
			{Adagideli}}, \ and\ \bibinfo {author} {\bibfnamefont {P.}~\bibnamefont
			{Jacquod}},\ }\href {\doibase 10.1103/PhysRevLett.98.196601} {\bibfield
		{journal} {\bibinfo  {journal} {Phys. Rev. Lett.}\ }\textbf {\bibinfo
			{volume} {98}},\ \bibinfo {pages} {196601} (\bibinfo {year}
		{2007})}\BibitemShut {NoStop}%
	\bibitem [{\citenamefont {Krich}\ and\ \citenamefont
		{Halperin}(2008)}]{Krich2008}%
	\BibitemOpen
	\bibfield  {author} {\bibinfo {author} {\bibfnamefont {J.~J.}\ \bibnamefont
			{Krich}}\ and\ \bibinfo {author} {\bibfnamefont {B.~I.}\ \bibnamefont
			{Halperin}},\ }\href {\doibase 10.1103/PhysRevB.78.035338} {\bibfield
		{journal} {\bibinfo  {journal} {Phys. Rev. B}\ }\textbf {\bibinfo {volume}
			{78}},\ \bibinfo {pages} {035338} (\bibinfo {year} {2008})}\BibitemShut
	{NoStop}%
	\bibitem [{\citenamefont {Adagideli}\ \emph {et~al.}(2010)\citenamefont
		{Adagideli}, \citenamefont {Jacquod}, \citenamefont {Scheid}, \citenamefont
		{Duckheim}, \citenamefont {Loss},\ and\ \citenamefont
		{Richter}}]{Adagideli2010}%
	\BibitemOpen
	\bibfield  {author} {\bibinfo {author} {\bibfnamefont {I.}~\bibnamefont
			{Adagideli}}, \bibinfo {author} {\bibfnamefont {P.}~\bibnamefont {Jacquod}},
		\bibinfo {author} {\bibfnamefont {M.}~\bibnamefont {Scheid}}, \bibinfo
		{author} {\bibfnamefont {M.}~\bibnamefont {Duckheim}}, \bibinfo {author}
		{\bibfnamefont {D.}~\bibnamefont {Loss}}, \ and\ \bibinfo {author}
		{\bibfnamefont {K.}~\bibnamefont {Richter}},\ }\href@noop {} {\bibfield
		{journal} {\bibinfo  {journal} {Phys. Rev. Lett.}\ }\textbf {\bibinfo
			{volume} {105}},\ \bibinfo {pages} {246807} (\bibinfo {year}
		{2010})}\BibitemShut {NoStop}%
	\bibitem [{\citenamefont {Stano}\ and\ \citenamefont
		{Jacquod}(2011)}]{Stano2011}%
	\BibitemOpen
	\bibfield  {author} {\bibinfo {author} {\bibfnamefont {P.}~\bibnamefont
			{Stano}}\ and\ \bibinfo {author} {\bibfnamefont {P.}~\bibnamefont
			{Jacquod}},\ }\href@noop {} {\bibfield  {journal} {\bibinfo  {journal} {Phys.
				Rev. Lett.}\ }\textbf {\bibinfo {volume} {106}},\ \bibinfo {pages} {206602}
		(\bibinfo {year} {2011})}\BibitemShut {NoStop}%
	\bibitem [{\citenamefont {Ramos}\ \emph {et~al.}(2018)\citenamefont {Ramos},
		\citenamefont {Vasconcelos},\ and\ \citenamefont {Barbosa}}]{Ramos2018}%
	\BibitemOpen
	\bibfield  {author} {\bibinfo {author} {\bibfnamefont {J.~G. G.~S.}\
			\bibnamefont {Ramos}}, \bibinfo {author} {\bibfnamefont {T.~C.}\ \bibnamefont
			{Vasconcelos}}, \ and\ \bibinfo {author} {\bibfnamefont {A.~L.~R.}\
			\bibnamefont {Barbosa}},\ }\href@noop {} {\bibfield  {journal} {\bibinfo
			{journal} {J. Appl. Phys.}\ }\textbf {\bibinfo {volume} {123}},\ \bibinfo
		{pages} {034304} (\bibinfo {year} {2018})}\BibitemShut {NoStop}%
	\bibitem [{\citenamefont {Taskinen}\ \emph {et~al.}(2008)\citenamefont
		{Taskinen}, \citenamefont {Starrett}, \citenamefont {Martin}, \citenamefont
		{Micolich}, \citenamefont {Hamilton}, \citenamefont {Simmons}, \citenamefont
		{Ritchie},\ and\ \citenamefont {Pepper}}]{Taskinen2008}%
	\BibitemOpen
	\bibfield  {author} {\bibinfo {author} {\bibfnamefont {L.~J.}\ \bibnamefont
			{Taskinen}}, \bibinfo {author} {\bibfnamefont {R.~P.}\ \bibnamefont
			{Starrett}}, \bibinfo {author} {\bibfnamefont {T.~P.}\ \bibnamefont
			{Martin}}, \bibinfo {author} {\bibfnamefont {A.}~\bibnamefont {Micolich}},
		\bibinfo {author} {\bibfnamefont {A.~R.}\ \bibnamefont {Hamilton}}, \bibinfo
		{author} {\bibfnamefont {M.~Y.}\ \bibnamefont {Simmons}}, \bibinfo {author}
		{\bibfnamefont {D.~A.}\ \bibnamefont {Ritchie}}, \ and\ \bibinfo {author}
		{\bibfnamefont {M.}~\bibnamefont {Pepper}},\ }\href@noop {} {\bibfield
		{journal} {\bibinfo  {journal} {Rev. Sci. Instrum.}\ }\textbf {\bibinfo
			{volume} {79}},\ \bibinfo {pages} {123901} (\bibinfo {year}
		{2008})}\BibitemShut {NoStop}%
	\bibitem [{\citenamefont {Benter}\ \emph {et~al.}(2013)\citenamefont {Benter},
		\citenamefont {Lehmann}, \citenamefont {Matsuyama}, \citenamefont {Hansen},
		\citenamefont {Heyn}, \citenamefont {Merkt},\ and\ \citenamefont
		{Jacob}}]{Benter2013}%
	\BibitemOpen
	\bibfield  {author} {\bibinfo {author} {\bibfnamefont {T.}~\bibnamefont
			{Benter}}, \bibinfo {author} {\bibfnamefont {H.}~\bibnamefont {Lehmann}},
		\bibinfo {author} {\bibfnamefont {T.}~\bibnamefont {Matsuyama}}, \bibinfo
		{author} {\bibfnamefont {W.}~\bibnamefont {Hansen}}, \bibinfo {author}
		{\bibfnamefont {C.}~\bibnamefont {Heyn}}, \bibinfo {author} {\bibfnamefont
			{U.}~\bibnamefont {Merkt}}, \ and\ \bibinfo {author} {\bibfnamefont
			{J.}~\bibnamefont {Jacob}},\ }\href@noop {} {\bibfield  {journal} {\bibinfo
			{journal} {Appl. Phys. Lett.}\ }\textbf {\bibinfo {volume} {102}},\ \bibinfo
		{pages} {212405} (\bibinfo {year} {2013})}\BibitemShut {NoStop}%
	\bibitem [{\citenamefont {Stano}\ \emph {et~al.}(2012)\citenamefont {Stano},
		\citenamefont {Fabian},\ and\ \citenamefont {Jacquod}}]{Stano2012}%
	\BibitemOpen
	\bibfield  {author} {\bibinfo {author} {\bibfnamefont {P.}~\bibnamefont
			{Stano}}, \bibinfo {author} {\bibfnamefont {J.}~\bibnamefont {Fabian}}, \
		and\ \bibinfo {author} {\bibfnamefont {P.}~\bibnamefont {Jacquod}},\
	}\href@noop {} {\bibfield  {journal} {\bibinfo  {journal} {Phys. Rev. B}\
		}\textbf {\bibinfo {volume} {85}},\ \bibinfo {pages} {241301(R)} (\bibinfo
		{year} {2012})}\BibitemShut {NoStop}%
	\bibitem [{\citenamefont {Buettiker}(1990)}]{Buettiker1990}%
	\BibitemOpen
	\bibfield  {author} {\bibinfo {author} {\bibfnamefont {M.}~\bibnamefont
			{B\"{u}ttiker}},\ }\href@noop {} {\bibfield  {journal} {\bibinfo  {journal}
			{Phys. Rev. B}\ }\textbf {\bibinfo {volume} {41}},\ \bibinfo {pages} {7906}
		(\bibinfo {year} {1990})}\BibitemShut {NoStop}%
	\bibitem [{Note1()}]{Note1}%
	\BibitemOpen
	\bibinfo {note} {See the Supplementary Material at [URL] for additional
		details on theory of spin-to-charge conversion and device characterization.}\BibitemShut {Stop}%
	\bibitem [{\citenamefont {Thomas}\ \emph {et~al.}(1996)\citenamefont {Thomas},
		\citenamefont {Nicholls}, \citenamefont {Simmons}, \citenamefont {Pepper},
		\citenamefont {Mace},\ and\ \citenamefont {Ritchie}}]{Thomas1996}%
	\BibitemOpen
	\bibfield  {author} {\bibinfo {author} {\bibfnamefont {K.~J.}~\bibnamefont
			{Thomas}}, \bibinfo {author} {\bibfnamefont {J.~T.}~\bibnamefont {Nicholls}},
		\bibinfo {author} {\bibfnamefont {M.~Y.}\ \bibnamefont {Simmons}}, \bibinfo
		{author} {\bibfnamefont {M.}~\bibnamefont {Pepper}}, \bibinfo {author}
		{\bibfnamefont {D.~R.}\ \bibnamefont {Mace}}, \ and\ \bibinfo {author}
		{\bibfnamefont {D.~A.}\ \bibnamefont {Ritchie}},\ }\href@noop {} {\bibfield
		{journal} {\bibinfo  {journal} {Phys. Rev. Lett.}\ }\textbf {\bibinfo
			{volume} {77}},\ \bibinfo {pages} {135} (\bibinfo {year} {1996})}\BibitemShut
	{NoStop}%
	\bibitem [{\citenamefont {Danneau}\ \emph {et~al.}(2008)\citenamefont
		{Danneau}, \citenamefont {Klochan}, \citenamefont {Clarke}, \citenamefont
		{Ho}, \citenamefont {Micolich}, \citenamefont {Simmons}, \citenamefont
		{Hamilton}, \citenamefont {Pepper},\ and\ \citenamefont
		{Ritchie}}]{Danneau2008}%
	\BibitemOpen
	\bibfield  {author} {\bibinfo {author} {\bibfnamefont {R.}~\bibnamefont
			{Danneau}}, \bibinfo {author} {\bibfnamefont {O.}~\bibnamefont {Klochan}},
		\bibinfo {author} {\bibfnamefont {W.~R.}\ \bibnamefont {Clarke}}, \bibinfo
		{author} {\bibfnamefont {L.~H.}\ \bibnamefont {Ho}}, \bibinfo {author}
		{\bibfnamefont {A.~P.}\ \bibnamefont {Micolich}}, \bibinfo {author}
		{\bibfnamefont {M.~Y.}\ \bibnamefont {Simmons}}, \bibinfo {author}
		{\bibfnamefont {A.~R.}\ \bibnamefont {Hamilton}}, \bibinfo {author}
		{\bibfnamefont {M.}~\bibnamefont {Pepper}}, \ and\ \bibinfo {author}
		{\bibfnamefont {D.~A.}\ \bibnamefont {Ritchie}},\ }\href@noop {} {\bibfield
		{journal} {\bibinfo  {journal} {Phys. Rev. Lett.}\ }\textbf {\bibinfo
			{volume} {100}},\ \bibinfo {pages} {016403} (\bibinfo {year}
		{2008})}\BibitemShut {NoStop}%
	\bibitem [{\citenamefont {Micolich}\ and\ \citenamefont
		{Z{\"{u}}licke}(2011)}]{Micolich2011}%
	\BibitemOpen
	\bibfield  {author} {\bibinfo {author} {\bibfnamefont {A.~P.}\ \bibnamefont
			{Micolich}}\ and\ \bibinfo {author} {\bibfnamefont {U.}~\bibnamefont
			{Z{\"{u}}licke}},\ }\href@noop {} {\bibfield  {journal} {\bibinfo  {journal}
			{J. Phys. Condens. Matter}\ }\textbf {\bibinfo {volume} {23}},\ \bibinfo
		{pages} {362201} (\bibinfo {year} {2011})}\BibitemShut {NoStop}%
	\bibitem [{\citenamefont {Iqbal}\ \emph {et~al.}(2013)\citenamefont {Iqbal},
		\citenamefont {Levy}, \citenamefont {Koop}, \citenamefont {Dekker},
		\citenamefont {de~Jong}, \citenamefont {van~der Velde}, \citenamefont
		{Reuter}, \citenamefont {Wieck}, \citenamefont {Aguado}, \citenamefont
		{Meir},\ and\ \citenamefont {van~der Wal}}]{Iqbal2013}%
	\BibitemOpen
	\bibfield  {author} {\bibinfo {author} {\bibfnamefont {M.~J.}\ \bibnamefont
			{Iqbal}}, \bibinfo {author} {\bibfnamefont {R.}~\bibnamefont {Levy}},
		\bibinfo {author} {\bibfnamefont {E.~J.}\ \bibnamefont {Koop}}, \bibinfo
		{author} {\bibfnamefont {J.~B.}\ \bibnamefont {Dekker}}, \bibinfo {author}
		{\bibfnamefont {J.~P.}\ \bibnamefont {de~Jong}}, \bibinfo {author}
		{\bibfnamefont {J.~H.~M.}\ \bibnamefont {van~der Velde}}, \bibinfo {author}
		{\bibfnamefont {D.}~\bibnamefont {Reuter}}, \bibinfo {author} {\bibfnamefont
			{A.~D.}\ \bibnamefont {Wieck}}, \bibinfo {author} {\bibfnamefont
			{R.}~\bibnamefont {Aguado}}, \bibinfo {author} {\bibfnamefont
			{Y.}~\bibnamefont {Meir}}, \ and\ \bibinfo {author} {\bibfnamefont {C.~H.}\
			\bibnamefont {van~der Wal}},\ }\href@noop {} {\bibfield  {journal} {\bibinfo
			{journal} {Nature}\ }\textbf {\bibinfo {volume} {501}},\ \bibinfo {pages}
		{79} (\bibinfo {year} {2013})}\BibitemShut {NoStop}%
	\bibitem [{\citenamefont {Bauer}\ \emph {et~al.}(2013)\citenamefont {Bauer},
		\citenamefont {Heyder}, \citenamefont {Schubert}, \citenamefont {Borowsky},
		\citenamefont {Taubert}, \citenamefont {Bruognolo}, \citenamefont {Schuh},
		\citenamefont {Wegscheider}, \citenamefont {von Delft},\ and\ \citenamefont
		{Ludwig}}]{Bauer2013}%
	\BibitemOpen
	\bibfield  {author} {\bibinfo {author} {\bibfnamefont {F.}~\bibnamefont
			{Bauer}}, \bibinfo {author} {\bibfnamefont {J.}~\bibnamefont {Heyder}},
		\bibinfo {author} {\bibfnamefont {E.}~\bibnamefont {Schubert}}, \bibinfo
		{author} {\bibfnamefont {D.}~\bibnamefont {Borowsky}}, \bibinfo {author}
		{\bibfnamefont {D.}~\bibnamefont {Taubert}}, \bibinfo {author} {\bibfnamefont
			{B.}~\bibnamefont {Bruognolo}}, \bibinfo {author} {\bibfnamefont
			{D.}~\bibnamefont {Schuh}}, \bibinfo {author} {\bibfnamefont
			{W.}~\bibnamefont {Wegscheider}}, \bibinfo {author} {\bibfnamefont
			{J.}~\bibnamefont {von Delft}}, \ and\ \bibinfo {author} {\bibfnamefont
			{S.}~\bibnamefont {Ludwig}},\ }\href@noop {} {\bibfield  {journal} {\bibinfo
			{journal} {Nature}\ }\textbf {\bibinfo {volume} {501}},\ \bibinfo {pages}
		{73} (\bibinfo {year} {2013})}\BibitemShut {NoStop}%
	\bibitem [{\citenamefont {Yeoh}\ \emph {et~al.}(2010)\citenamefont {Yeoh},
		\citenamefont {Srinivasan}, \citenamefont {Martin}, \citenamefont {Klochan},
		\citenamefont {Micolich},\ and\ \citenamefont {Hamilton}}]{Yeoh2010}%
	\BibitemOpen
	\bibfield  {author} {\bibinfo {author} {\bibfnamefont {L.~A.}\ \bibnamefont
			{Yeoh}}, \bibinfo {author} {\bibfnamefont {A.}~\bibnamefont {Srinivasan}},
		\bibinfo {author} {\bibfnamefont {T.~P.}\ \bibnamefont {Martin}}, \bibinfo
		{author} {\bibfnamefont {O.}~\bibnamefont {Klochan}}, \bibinfo {author}
		{\bibfnamefont {A.~P.}\ \bibnamefont {Micolich}}, \ and\ \bibinfo {author}
		{\bibfnamefont {A.~R.}\ \bibnamefont {Hamilton}},\ }\href@noop {} {\bibfield
		{journal} {\bibinfo  {journal} {Rev. Sci. Instrum.}\ }\textbf {\bibinfo
			{volume} {81}},\ \bibinfo {pages} {113905} (\bibinfo {year}
		{2010})}\BibitemShut {NoStop}%
		\bibitem [{Note2()}]{Note2}%
	\BibitemOpen
	\bibinfo {note} {The fitting ranges $|B| \leq 0.5, 0.75, 1, 1.25$ T give the
		same peak positions (see Sec.~S5 of the Supplementary Material \cite
		{Note2}).}\BibitemShut {Stop}%
	\bibitem [{\citenamefont {Jungwirth}\ \emph {et~al.}(2012)\citenamefont
		{Jungwirth}, \citenamefont {Wunderlich},\ and\ \citenamefont
		{Olejn{\'{i}}k}}]{Jungwirth2012}%
	\BibitemOpen
	\bibfield  {author} {\bibinfo {author} {\bibfnamefont {T.}~\bibnamefont
			{Jungwirth}}, \bibinfo {author} {\bibfnamefont {J.}~\bibnamefont
			{Wunderlich}}, \ and\ \bibinfo {author} {\bibfnamefont {K.}~\bibnamefont
			{Olejn{\'{i}}k}},\ }\href {http://dx.doi.org/10.1038/nmat3279} {\bibfield
		{journal} {\bibinfo  {journal} {Nat. Mater.}\ }\textbf {\bibinfo {volume}
			{11}},\ \bibinfo {pages} {382} (\bibinfo {year} {2012})}\BibitemShut
	{NoStop}%
	\bibitem [{\citenamefont {Balakrishnan}\ \emph {et~al.}(2014)\citenamefont
		{Balakrishnan}, \citenamefont {Koon}, \citenamefont {Avsar}, \citenamefont
		{Ho}, \citenamefont {Lee}, \citenamefont {Jaiswal}, \citenamefont {Baeck},
		\citenamefont {Ahn}, \citenamefont {Ferreira}, \citenamefont {Cazalilla},
		\citenamefont {{Castro Neto}},\ and\ \citenamefont
		{{\"{O}}zyilmaz}}]{Balakrishnan2014a}%
	\BibitemOpen
	\bibfield  {author} {\bibinfo {author} {\bibfnamefont {J.}~\bibnamefont
			{Balakrishnan}}, \bibinfo {author} {\bibfnamefont {G.~K.~W.}\ \bibnamefont
			{Koon}}, \bibinfo {author} {\bibfnamefont {A.}~\bibnamefont {Avsar}},
		\bibinfo {author} {\bibfnamefont {Y.}~\bibnamefont {Ho}}, \bibinfo {author}
		{\bibfnamefont {J.~H.}\ \bibnamefont {Lee}}, \bibinfo {author} {\bibfnamefont
			{M.}~\bibnamefont {Jaiswal}}, \bibinfo {author} {\bibfnamefont {S.-J.}\
			\bibnamefont {Baeck}}, \bibinfo {author} {\bibfnamefont {J.-H.}\ \bibnamefont
			{Ahn}}, \bibinfo {author} {\bibfnamefont {A.}~\bibnamefont {Ferreira}},
		\bibinfo {author} {\bibfnamefont {M.~a.}\ \bibnamefont {Cazalilla}}, \bibinfo
		{author} {\bibfnamefont {A.~H.}\ \bibnamefont {{Castro Neto}}}, \ and\
		\bibinfo {author} {\bibfnamefont {B.}~\bibnamefont {{\"{O}}zyilmaz}},\
	}\href@noop {} {\bibfield  {journal} {\bibinfo  {journal} {Nat. Commun.}\
		}\textbf {\bibinfo {volume} {5}},\ \bibinfo {pages} {4748} (\bibinfo {year}
		{2014})}\BibitemShut {NoStop}%
	\bibitem [{\citenamefont {Sinova}\ \emph {et~al.}(2015)\citenamefont {Sinova},
		\citenamefont {Valenzuela}, \citenamefont {Wunderlich}, \citenamefont
		{Back},\ and\ \citenamefont {Jungwirth}}]{Sinova2015}%
	\BibitemOpen
	\bibfield  {author} {\bibinfo {author} {\bibfnamefont {J.}~\bibnamefont
			{Sinova}}, \bibinfo {author} {\bibfnamefont {S.~O.}\ \bibnamefont
			{Valenzuela}}, \bibinfo {author} {\bibfnamefont {J.}~\bibnamefont
			{Wunderlich}}, \bibinfo {author} {\bibfnamefont {C.~H.}\ \bibnamefont
			{Back}}, \ and\ \bibinfo {author} {\bibfnamefont {T.}~\bibnamefont
			{Jungwirth}},\ }\href {\doibase 10.1103/RevModPhys.87.1213} {\bibfield
		{journal} {\bibinfo  {journal} {Rev. Mod. Phys.}\ }\textbf {\bibinfo {volume}
			{87}},\ \bibinfo {pages} {1213} (\bibinfo {year} {2015})}\BibitemShut
	{NoStop}%
	\bibitem [{\citenamefont {Appleyard}\ \emph {et~al.}(1998)\citenamefont
		{Appleyard}, \citenamefont {Nicholls}, \citenamefont {Simmons}, \citenamefont
		{Tribe},\ and\ \citenamefont {Pepper}}]{Appleyard1998}%
	\BibitemOpen
	\bibfield  {author} {\bibinfo {author} {\bibfnamefont {N.~J.}\ \bibnamefont
			{Appleyard}}, \bibinfo {author} {\bibfnamefont {J.~T.}\ \bibnamefont
			{Nicholls}}, \bibinfo {author} {\bibfnamefont {M.~Y.}\ \bibnamefont
			{Simmons}}, \bibinfo {author} {\bibfnamefont {W.~R.}\ \bibnamefont {Tribe}},
		\ and\ \bibinfo {author} {\bibfnamefont {M.}~\bibnamefont {Pepper}},\
	}\href@noop {} {\bibfield  {journal} {\bibinfo  {journal} {Phys. Rev. Lett.}\
		}\textbf {\bibinfo {volume} {81}},\ \bibinfo {pages} {3491} (\bibinfo {year}
		{1998})}\BibitemShut {NoStop}%
	\bibitem [{\citenamefont {Bakker}\ \emph {et~al.}(2010)\citenamefont {Bakker},
		\citenamefont {Slachter}, \citenamefont {Adam},\ and\ \citenamefont {{Van
				Wees}}}]{Bakker2010}%
	\BibitemOpen
	\bibfield  {author} {\bibinfo {author} {\bibfnamefont {F.~L.}\ \bibnamefont
			{Bakker}}, \bibinfo {author} {\bibfnamefont {A.}~\bibnamefont {Slachter}},
		\bibinfo {author} {\bibfnamefont {J.~P.}\ \bibnamefont {Adam}}, \ and\
		\bibinfo {author} {\bibfnamefont {B.~J.}\ \bibnamefont {{van Wees}}},\
	}\href@noop {} {\bibfield  {journal} {\bibinfo  {journal} {Phys. Rev. Lett.}\
		}\textbf {\bibinfo {volume} {105}},\ \bibinfo {pages} {136601} (\bibinfo
		{year} {2010})}\BibitemShut {NoStop}%
	\bibitem [{Note3()}]{Note3}%
	\BibitemOpen
	\bibinfo {note} {As an example, the linear spin-to-charge measurements in Fig.~\ref{fig:linear_regime_spin_currents_old} took $\sim30$ hours, as they needed data at both $+B$ and $-B$. The equivalent 	non-linear spin-to-charge conversion, which only involved measuring $V_3$ as a function of $V_{\mathrm{qpc}}$ at $B=0$, took only $\sim 10$ minutes. }
	\BibitemShut {NoStop}%
\end{thebibliography}%

%

\end{document}


\title{Supplement of ``A non-linear spin filter for non-magnetic materials at zero magnetic field"}
\author{E. Marcellina}
\altaffiliation{Present address: School of Physical and Mathematical Sciences, Nanyang Technological University, 21 Nanyang Link, Singapore 637371. Email: emarcellina@ntu.edu.sg}
\affiliation{School of Physics, The University of New South Wales, Sydney, Australia}

\author{A. Srinivasan}
\affiliation{School of Physics, The University of New South Wales, Sydney, Australia}

\author{F. Nichele}
\affiliation{IBM Research - Zurich,	S\"{a}umerstrasse 4, 8803 R\"{u}schlikon, Switzerland}

\author{P. Stano}
\affiliation{RIKEN Center for Emergent Matter Science (CEMS), Wako, Saitama 351-0198, Japan}
\affiliation{Department of Applied Physics, School of Engineering, University of Tokyo, 7-3-1 Hongo, Bunkyo-ku, Tokyo 113-8656, Japan}
\affiliation{Institute of Physics, Slovak Academy of Sciences, 845 11 Bratislava, Slovakia}

\author{D. A. Ritchie}
\affiliation{Cavendish Laboratory, University of Cambridge, J. J. Thomson Avenue, Cambridge CB3 0HE, United Kingdom}

\author{I. Farrer}
\affiliation{Cavendish Laboratory, University of Cambridge, J. J. Thomson Avenue, Cambridge CB3 0HE, United Kingdom}

\author{Dimitrie Culcer}
\affiliation{School of Physics, The University of New South Wales, Sydney, Australia}

\author{A. R. Hamilton}
\affiliation{School of Physics, The University of New South Wales, Sydney, Australia}
\maketitle

We provide a theoretical background for spin-to-charge conversion in ballistic systems in Sec.~\ref{sec:theory_detail}. We continue with the characterization of the two-dimensional hole system in Sec.~\ref{sec:2D_char}. We describe the quantum point contact used as the energy-selective barrier in Sec.~\ref{sec:QPC_char} and the drive channel in Sec.~\ref{sec:drive_char}. We finally discuss the effect of the in-plane magnetic field on the linear spin signal in Sec.~\ref{sec:linear_fit_detail} and on the non-linear spin signal in  Sec.~\ref{sec:non_linear_in_plane_B}.

\section{Theory}
\label{sec:theory_detail}

\subsection{Landauer-B\"uttiker formalism}

In a multi-terminal mesoscopic structure, the current $I_{\mathrm{qpc}}$ across a QPC between terminals $a$ and $b$ can be derived using the Landauer-B\"uttiker formula, which reads \cite{Stano2012}:
\begin{equation}
I_{\mathrm{qpc}} = G_1 e \delta V + G_2 \left(\delta{\mu}_{a}^2 - \delta{\mu}_{b}^2\right) + G_3 \left(\delta \mu_{a} - \delta \mu_{b} \right)  + G_4 \left(\delta \mu_{a} + \delta \mu_{b}\right) e \delta V,
\label{eq:current}
\end{equation}
where $\delta V = V_a - V_b$, with $V_{i}$ ($\delta \mu_{i}$) the electric potential (spin accumulation in units of energy, defined as the chemical potential difference of spin up and down species) at terminal $i \in \{ a, b\}$. The conductances are 
\begin{subequations}
	\label{eq:conductances}
	\begin{align}
	G_1 & = \frac{2e}{h} \int dE \left(-\partial_E f(E) \right) T(E, B), \label{subeq:conductances}\\ 
	G_2 & = \frac{e}{h} \int dE \left(-\partial_E f(E) \right) \left[\partial_E T(E, B)\right],\\ 
	G_3 & = \frac{2e}{h} \int dE \left(-\partial_E f(E) \right) \delta T(E, B), \\
	G_4 & = \frac{e}{h} \int dE \left(-\partial_E f(E) \right) \left[\partial_E \delta T(E, B)\right] 
	\end{align}
\end{subequations}
where $f(E)$ is the Fermi-Dirac distribution, $T(E, B)$ is the transmission probability of the energy barrier, and $\delta T(E, B)$ accounts for the dependence of the transmission probability between terminal $i$ and terminal $j$ on the spin orientation $\sigma$ ($\sigma = \pm1$ for opposite spin orientations), i.e. $T_{ji}^{\sigma\sigma} = T(E, B) + \sigma \delta T(E, B)$. Equations \eqref{eq:current} and \eqref{eq:conductances} are valid in the limit of low excitation current such that $f(E-\mu_a) - f(E-\mu_b) \propto \partial_E f(E)$, where $\mu_{i}$ is the chemical potential in terminal $i$. For a QPC, the transmission is \cite{Buettiker1990}
\begin{equation}
T(E) = \{1 + \exp\left[-2\pi (E - e \alpha_{\mathrm{qpc}} V_{\mathrm{qpc}} - \sigma \mu B)/ \hbar \Omega_{\mathrm{qpc}} \right]\}^{-1}, 
\label{eq:QPC_transmission_probability}
\end{equation}
where $\alpha_{\mathrm{qpc}}$ is the QPC lever arm, $V_{\mathrm{qpc}} $ is the QPC gate voltage, $\mu = (g/2) \mu_B$, with $g$ being the $g-$factor, $\mu_B$ is the Bohr magneton, $B$ is the magnetic field, and $h\Omega_{\mathrm{qpc}}$ is the curvature of the saddle potential. In our sample, spin accumulation is only generated in the driving channel, thus we set $\delta \mu_b = 0$ and $\delta \mu_{a} \equiv \delta \mu_{s}$ throughout this work.

The four conductance terms in Eq.~\eqref{eq:conductances} above have the following interpretations. The first term $G_1$ is the (usual) charge conductance through a QPC. The second term $G_2$ is the non-linear spin-to-charge coupling term. Note that the first two terms in Eq.~\eqref{eq:current} are insensitive to the sign of spin accumulation, as they are only dependent on $T(E,B)$. The third and fourth terms are the linear spin-to-charge coupling terms which are spin sensitive (as both depend on $ \delta T(E,B) \approx - \sigma \mu  B \partial_E T(E,B)$), and, for the transmission probability in Eq.~\eqref{eq:QPC_transmission_probability}, are finite only in the presence of a magnetic field $B$. Thus, $G_2$ is symmetric in $B$ whereas $G_3$ and $G_4$ are antisymmetric in $B$. Furthermore, Eq.~\eqref{eq:conductances} predicts that both $G_2$ and $G_3$ are proportional to the transconductance, whereas $G_4$ is proportional to the derivative of the transconductance with respect to the QPC voltage.

\subsection{Relation between the three-terminal voltage $V_3$ and spin accumulation $\delta \mu_s$}

We now decompose the term $\delta V$ in Eq.~\eqref{eq:current} in terms of its harmonics and relate each component to the three-terminal voltage $V_3$. We assume a sinusoidal excitation 	$I_{\mathrm{sd}}(t)  = I_{\mathrm{sd}} \sin(\omega t)$, so that:
\begin{subequations}
	\label{eq:signals_time_dependence}
	\begin{align}
	\delta \mu_s (t) & = \delta \mu_s \sin(\omega t),\\ 
	\delta V (t) & = \delta V(\omega) \sin(\omega t) +  \delta V(2 \omega) \sin(2 \omega t + \phi),
	\end{align}
\end{subequations}
where $\phi$ is the phase angle of the second harmonic component of $\delta V(t)$.

\subsubsection{Linear regime}

Setting $I_{\mathrm{qpc}} = 0$, and collecting terms oscillating at frequency $\omega$, Eq.~\eqref{eq:current} reads:
\begin{equation}
	-G_1 e \delta V(\omega) = (G_3 + G_4  e \delta V(\omega)) \delta \mu_{s},
	\label{eq:linear_V3_peak}
\end{equation}
where $\delta \mu_s$ is the spin accumulation in the drive channel. At $B \rightarrow 0$, at the point where the QPC transconductance is maximal, i.e. where $E = E_F = e \alpha_{\mathrm{qpc}} V_{\mathrm{qpc}}$, $G_4 = 0$ and Eq.~\eqref{eq:linear_V3_peak} becomes
\begin{equation}
	\begin{array}[b]{rcl}
	- G_1(E = E_F) e \delta V(\omega) &= & G_3 (E = E_F) \delta \mu_{s} \\
	(1/2) e \delta V(\omega)  &=& (g/2) \mu_B B \frac{\pi}{2 \hbar \Omega_{\mathrm{qpc}}} \delta \mu_s.
	\end{array}
\end{equation}
Or equivalently, 
\begin{equation}
	\frac{\partial (\delta V(\omega))}{\partial B}\Bigr|_{{B}=0} = 	\frac{\partial V_3(\omega)}{\partial B}\Bigr|_{{B}=0} = \frac{g \mu_B  \pi}{2 e \hbar \Omega_{\mathrm{qpc}}} \delta \mu_s.
	\label{eq:linear_spin_accumulation}
\end{equation}
Eq.~\eqref{eq:linear_spin_accumulation} explains the result of Fig.~2h, i.e. in the limit of small excitation current $I_{\mathrm{sd}}$, the signal derivative  $\frac{\partial V_3(\omega)}{\partial B}\Bigr|_{{B}=0}$ is proportional to $\delta \mu_s$ and hence $I_{sd}$.

\subsubsection{Non-linear regime}

Again, setting $I_{\mathrm{qpc}} = 0$, and collecting terms oscillating at frequency $2\omega$, and neglecting $G_4$, Eq.~\eqref{eq:current} reads:
\begin{equation}
-G_1 e \delta V(2\omega) =  (1/2) G_2  (\delta \mu_{s})^2.
\label{eq:non_linear_V3_peak}
\end{equation}
At $B = 0$, at the point where the QPC transconductance is maximal, that is where $E = E_F = e \alpha_{\mathrm{qpc}} V_{\mathrm{qpc}}$, Eq.~\eqref{eq:non_linear_V3_peak} becomes
\begin{equation}
\begin{array}[b]{rcl}
- G_1(E = E_F) e \delta V(2\omega) &= & (1/2) G_2 (E = E_F)  \delta \mu_{s}^2 \\
 \delta V(2\omega)  &=& -(1/2) \frac{\pi}{e\hbar \Omega_{\mathrm{qpc}}}\delta \mu_s^2 .
\end{array}
\label{eq:non_linear_spin_accumulation}
\end{equation}
Eq.~\eqref{eq:non_linear_spin_accumulation} explains the result of Fig.~3c, i.e. in the limit of small $I_{\mathrm{sd}}$, the non-linear three-terminal voltage $
\delta V(2\omega) = V_3(2\omega)$ is proportional to $\delta \mu_s^2$ and hence $I_{sd}^2$.

\begin{figure}
	\begin{center}
		\includegraphics[scale=0.7]{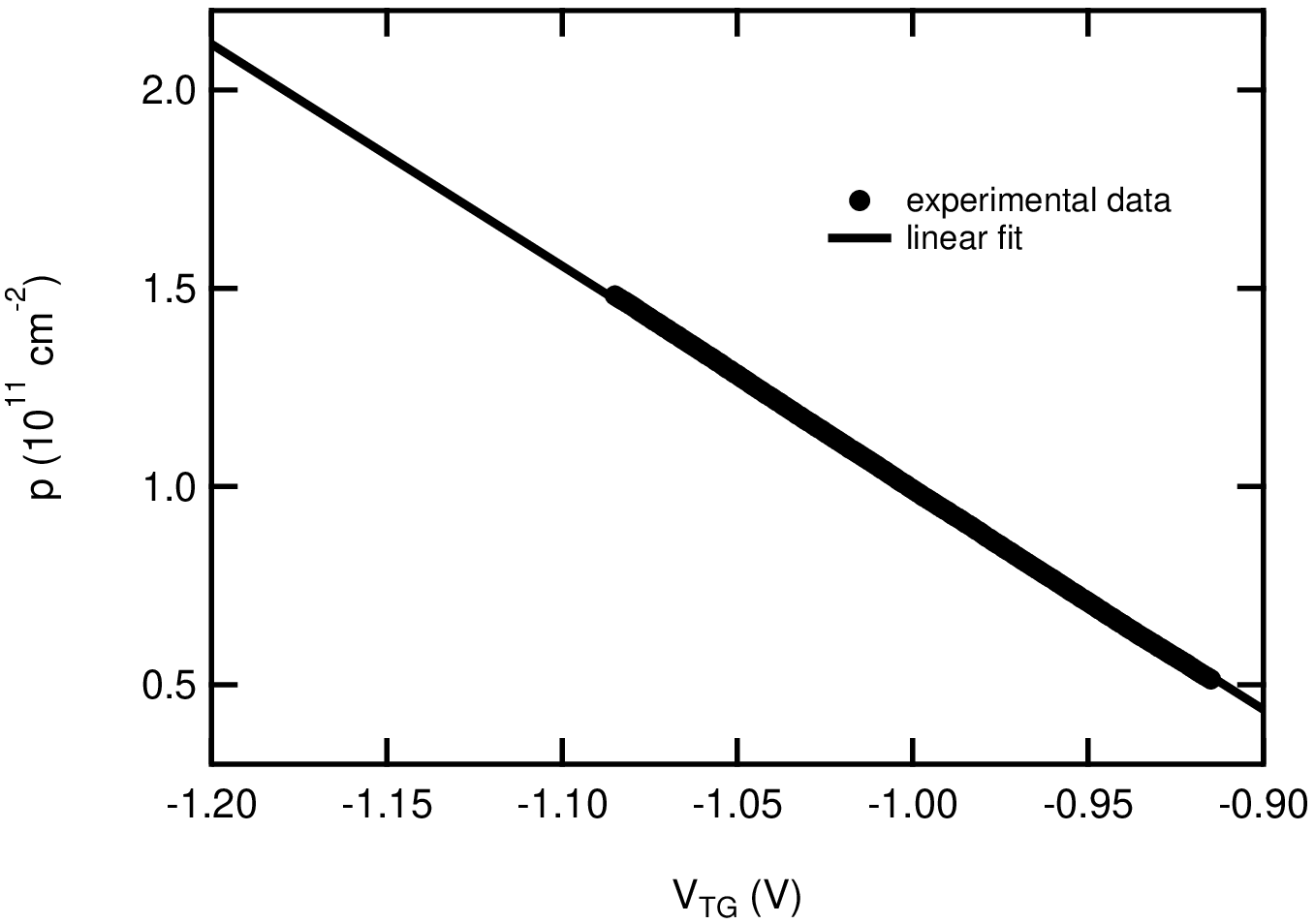}
		\caption{\label{fig:V_TG_vs_dens}The two-dimensional (2D) hole density versus top gate voltage, showing that the density	ranges from $4.4 \times 10^{10}$ cm$^{-2}$ to $2.1 \times 10^{11}$ cm$^{-2}$. Measurements were done in the Oxford Instruments K100 dilution refrigerator at the base temperature.}
	\end{center}
\end{figure}

\section{Characterization of the Two-dimensional Hole Gas}
\label{sec:2D_char}
 
\begin{figure}
	\begin{center}
		\includegraphics[scale=0.7]{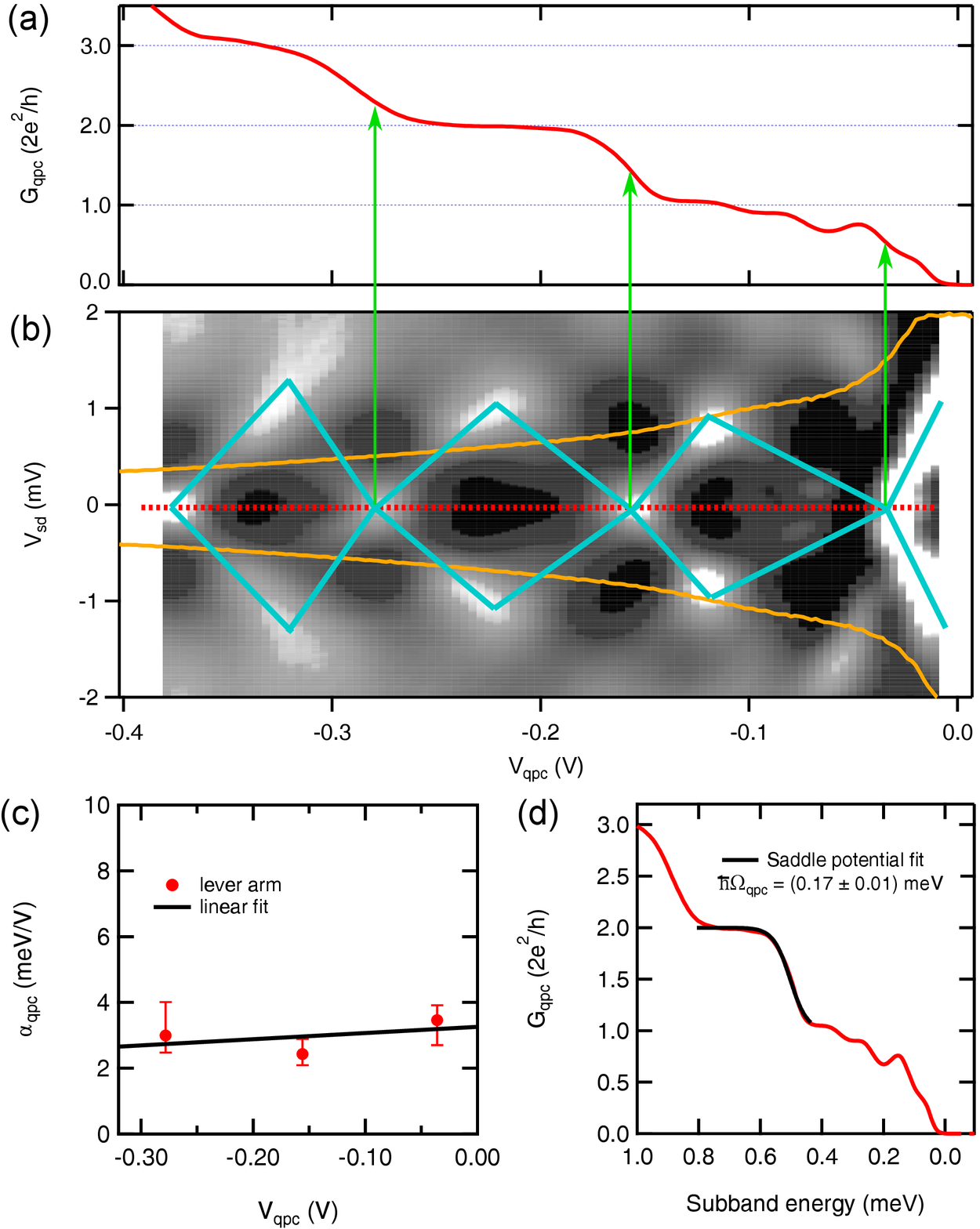}
		\caption{(a) The conductance $G_{\mathrm{qpc}}$ of the QPC as a function of the QPC gate voltage $V_{\mathrm{qpc}}$. (b) A greyscale plot of the transconductance against the source-drain voltage $V_{\mathrm{SD}}$ as a function of $V_{\mathrm{qpc}}$. Here, light (dark) shades correspond to transconductance maxima	(minima). The blue lines depict the one-dimensional subband edges whereas the orange lines denote the voltage drop across the QPC when a two-terminal DC voltage of $\pm 2$ mV is applied. (c) The lever arm $\alpha_{\mathrm{qpc}}$ for the QPC side gate, which serves as the conversion factor between the subband energy spacing and	the side gate voltage. For the first subband, $\alpha_{\mathrm{qpc}} = 3.46$ meV/V whilst for the second subband $\alpha_{\mathrm{qpc}} = 2.43$ meV/V. (d) The QPC conductance in (a) plotted against the subband energy. A saddle potential fit (Eq.~\eqref{eq:QPC_transmission_probability}) for $B = 0$ T reveals a curvature of  $\hbar \Omega_{\mathrm{qpc}} = 0.27 \pm 0.02$ meV for the second subband.}
		\label{fig:qpc_char}
	\end{center}
\end{figure}

\begin{figure}
	\begin{center}
		\includegraphics[scale=0.7]{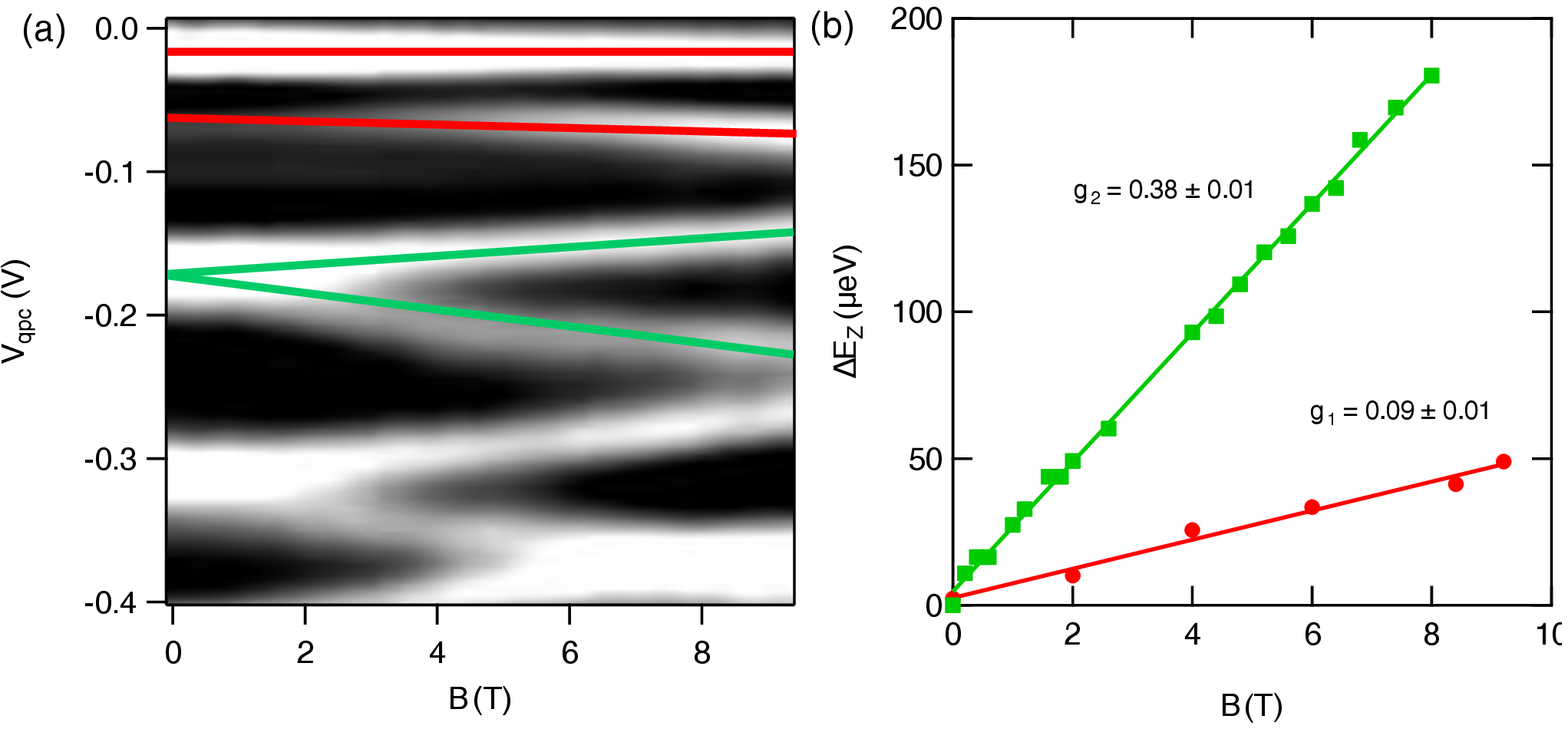}
		\caption{(a) A greyscale plot of the transconductance as a function of $B$. The red and green lines track the Zeeman splitting. (b) The Zeeman energy splitting $\Delta E_Z$ versus the in-plane magnetic field $B$, giving a $g-$factor $g_1 = 0.09 \pm 0.01$ and $g = g_2 = 0.38 \pm 0.01$ for the first and second subbands, respectively. Here the Zeeman energy is obtained by multiplying the corresponding side gate voltage by the lever arm $\alpha_{\mathrm{qpc}}$ from Fig.~\ref{fig:qpc_char}c.}
		\label{fig:zeeman_char}
	\end{center}
\end{figure} 

All measurements reported in this work were performed in a dilution fridge with a piezorotator \cite{Yeoh2010}. Prior to performing the spin-to-charge conversion measurements, we first characterized the two-dimensional hole density and mobility. These characterization measurements were performed by making the conduction channels two-dimensional, which was accomplished by setting the side gate voltages of the device in Fig.~2a to -0.81 V. Fig.~\ref{fig:V_TG_vs_dens} shows that the two-dimensional hole density that can be induced in our device ranges from $4.4 \times 10^{10}$ cm$^{-2}$ to $2.1 \times 10^{11}$ cm$^{-2}$. For the measurements presented in the main text, the top gate was maintained at $-1.18$ V, which corresponds to a two-dimensional hole density of $p = 2\times 10^{11}$ cm$^{-2}$ and a Fermi wavelength of $\lambda_F = 56$ nm. Separate measurements on another two-dimensional hole device made from the same heterostructure revealed a mobility of $\mu = 550,000$ cm$^{2}$V$^{-1}$s$^{-1}$ for $p = 2\times 10^{11}$ cm$^{-2}$. The mean free path $l$ calculated using $l = \frac{\hbar k_F \mu}{e}$ gives $l \approx 4$ $\mu$m, which means that the charge transport is in the ballistic regime.

\subsection*{Description of the spin-orbit interaction in our device}

We now characterize the spin-orbit interaction in our device, using our previous work in  Ref.~\cite{Marcellina2017}. Since our device is a heterojunction where the holes are confined in a quasi-triangular well, a Rashba spin-orbit interaction of the form $H_{R} = i \alpha_R (k_-^3 \sigma_+ - k_+^3 \sigma_-)$ is present. For a density of $p = 2\times 10^{11}$ cm$^{-2}$, the Rashba spin splitting is found to be  $\frac{p_+ - p_-}{p_+ + p_-} \approx 30 \%$, where $p_{\pm}$ are the spin-split densities, and the Rashba spin-orbit energy splitting is $\Delta_{\mathrm{SO}} = 1.73$ meV. The spin-split effective masses of holes in heterojunctions are different ($m_+ =$ 0.36 $m_0$ and $m_- = $0.11 $m_0$, where $m_{\pm}$ are the effective masses of the $p_{\pm}$ subbands), and give a density-averaged value of $m^* \equiv \frac{m_+ p_+ + m_- p_-}{p_+ + p_-} = 0.28~m_0$. The spin-orbit precession length is given as $l_{\mathrm{SO}} \equiv v_F \tau_{\mathrm{SO}} = (\hbar k_F/m^*)(2\hbar /  \Delta_{\mathrm{SO}})$ = 35 nm, where $v_F$ is the Fermi velocity and $\tau_{\mathrm{SO}}$ is the spin-orbit precession time. Finally, we note that the Dresselhaus spin-orbit interaction is negligible on our system \cite{Marcellina2017}.
 
\section{QPC characterization}
\label{sec:QPC_char}

Here we report four-terminal characterization results for the QPC in the middle of the device (see Figs. 2a and 2b of the main text). To this end, we set the surface gate voltage to 0 V, so that one-dimensional conduction channels form a `K'-shape (Fig.~1a of the main text). Since the dimensions of the `K-bar' are less than 4 $\mu$m, the transport through the `K-bar' is ballistic. 

Fig.~\ref{fig:qpc_char}a shows clear conductance plateaus for the first three one-dimensional subbands. There is a kink at $G = 0.7 \times 2e^2/h$ due to the well-known ``0.7 feature", whose origin is still elusive  \cite{Thomas1996,Danneau2008,Micolich2011,Iqbal2013,Bauer2013,Kawamura2015}. However, to simplify the analysis, here we assume that the QPC transmission probability obeys Eq.~\eqref{eq:QPC_transmission_probability}. 

Source-drain biasing measurements (Fig.~\ref{fig:qpc_char}b) yield the lever arm $\alpha_{\mathrm{qpc}}$, i.e. the conversion between the QPC gate voltage and the subband energy spacings. The bright (dark) regions in the grayscale map correspond to the transconductance maxima (minima) (see also the green arrows highlighting the connection between the transconductance maxima and the risers in the conductance). The orange lines denote the actual voltage drop across the QPC when a two-terminal DC voltage of $\pm 2$ mV is applied. The lever arm $\alpha_{\mathrm{qpc}} = \partial V_{\mathrm{SD}}/\partial {V_{\mathrm{qpc}}}$ for the first subband is found to be $3.46$ meV/V and $2.43$ meV/V (Fig.~\ref{fig:qpc_char}c). Once the lever arm is known, the conductance can be plotted as a function of one-dimensional subband energy instead of side gate voltage (Fig.~\ref{fig:qpc_char}d). Fitting the result with the transmission probability (Eq.~\eqref{eq:QPC_transmission_probability} with $B = 0$ T) we find that the saddle potential curvature is $\hbar \Omega_{\mathrm{qpc}} = 0.17 \pm 0.01$ meV for the second subband. We note that the parameter $\hbar \Omega_{\mathrm{qpc}}$ sets the upper temperature limit for spin-to-charge conversion to happen: below temperature $T = \frac{\hbar \Omega_{\mathrm{qpc}}}{k_B} \approx 2$ K, where $k_B$ is the Boltzmann constant, spin-to-charge conversion is expected to occur in our device. 
 
Fig.~\ref{fig:zeeman_char}a shows a color map of the QPC transconductance as a function of the QPC voltage $V_{\mathrm{qpc}}$ and $B$, where the solid lines track the Zeeman splitting. From Fig.~\ref{fig:zeeman_char}a and source-drain biasing measurements (Figs. \ref{fig:qpc_char}a-c), one can extract the Zeeman energy $\Delta E_Z = \alpha_{\mathrm{qpc}} V_{\mathrm{qpc}}$ \cite{Patel1991} as a function of $B$ (Fig.~\ref{fig:zeeman_char}b). We find that the $g-$factor is $g_{1} = 0.09 \pm 0.01$ and $g \equiv g_{2} = 0.38 \pm 0.01$ for the first and second subbands, respectively. 

\section{Driving channel characterization}
\label{sec:drive_char}

\begin{figure}
	 	\begin{center}
		\includegraphics[scale = 0.75,trim={0cm 0cm 0cm 0cm}]{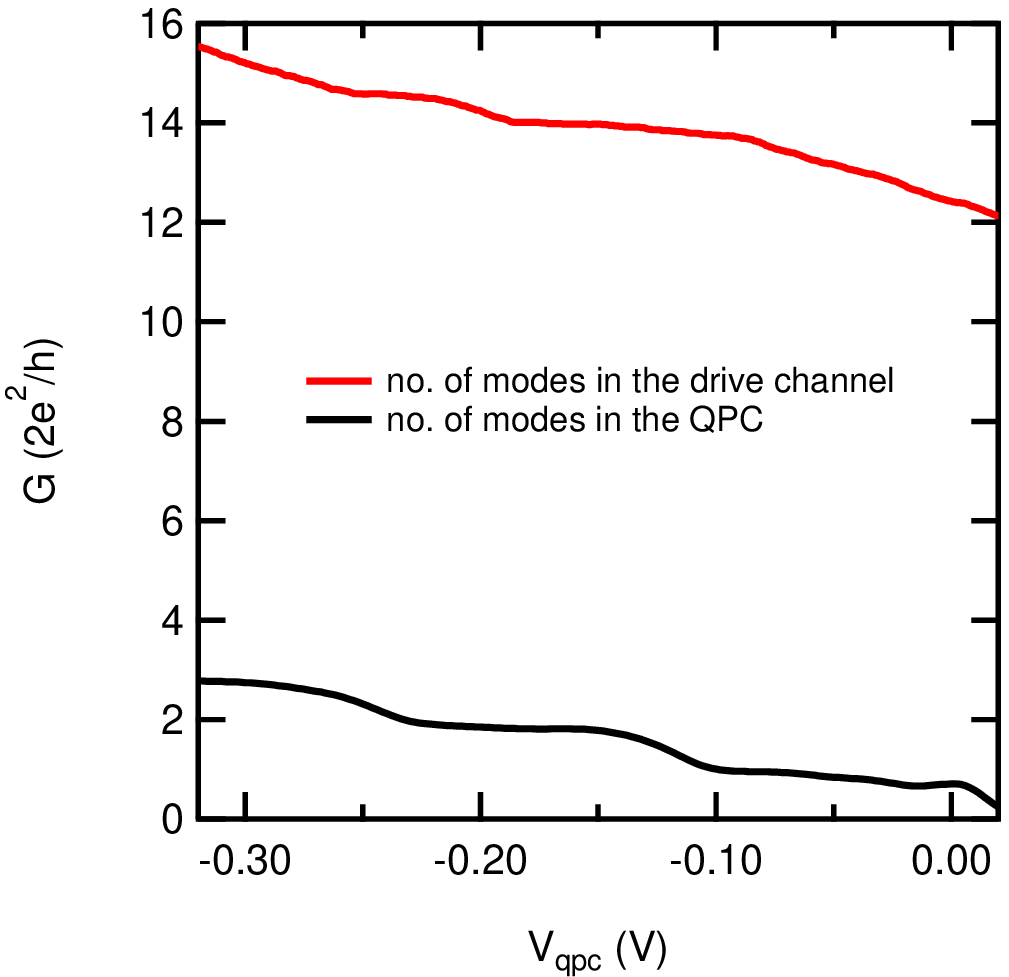}
		\caption{The number of modes in the QPC and in the drive channel at $B = 0$ T as a function of the QPC gate voltage $V_{\mathrm{qpc}}$, showing that $N_{\mathrm{drive}} \gg  N_{\mathrm{qpc}}$.}
		\label{fig:leg_char}
	\end{center}
\end{figure}
 
Using the two-terminal voltage in Fig.~2c for $B = 0$ T, and estimating a series resistance $R_s$ of 11.4 k$\mathrm{\Omega}$, the conductance of the charge conduction channel is found using the relation $G_C = \frac{h}{2e^2} \times \frac{1}{V_{\mathrm{d}}/I_{\mathrm{sd}} - R_s}$. The results are shown in Fig.~\ref{fig:leg_char}. At the QPC voltage where the correlation between the spin current and the transconductance is strongest (Fig.~2g), i.e. at the second QPC riser, there are 14 modes in the drive channel, which fulfills the requirement $N_{\mathrm{drive}} \gg N_{\mathrm{qpc}}$. We use $N_{\mathrm{drive}}$ to calculate the spin Hall angle $\Theta$, and find it to be 6.8\%. 

For comparison, the typical spin Hall angles range from less than 1\% to over 30\%. For instance, the spin Hall angle was found to be 0.15\% in $n-$GaAs \cite{Olejnik2012}, 34\% in $p-$GaAs \cite{Nichele2015}, 8.5\% in Pt \cite{Ganguly2014}, and 8.4\% in Au \cite{Wang2014}.   

\section{Linear fit of the first-harmonic three-terminal voltage as a function of $B$}
\label{sec:linear_fit_detail}

\begin{figure}[ht]
	\begin{center}
		\includegraphics[scale = 0.7,trim={0cm 0cm 0cm 0cm}]{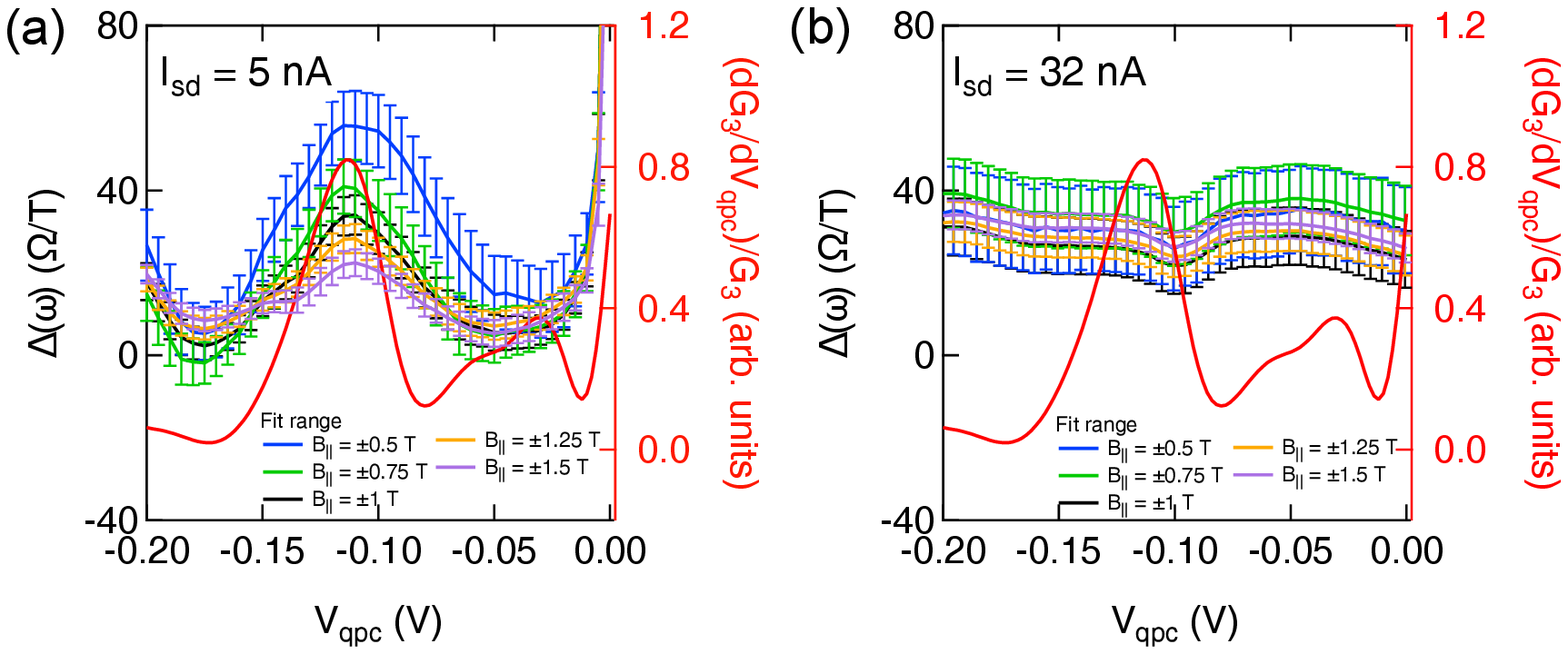}
		\caption{The asymmetry $\Delta(\omega)$ in the linear response regime, plotted for different fitting ranges for (a) $I_{\mathrm{sd}} = 5$ nA and (b) $I_{\mathrm{sd}} = 32$ nA. Panel (a) shows that the locations of the extrema of $\Delta(\omega)$ are independent of the fitting range. This means that our results agree with the theoretical prediction that the linear spin signal is proportional to the QPC transconductance. Panel (b) shows that at $I_{\mathrm{sd}} = 32$ nA the spin signal is suppressed.}
		\label{fig:spin_currents_different_ranges}
	\end{center}
\end{figure}

Here we examine the relation between the linear spin current and the in-plane magnetic field for different excitation currents of $I_{\mathrm{sd}} = 5$ nA and 32 nA. Fig.~\ref{fig:spin_currents_different_ranges} shows the asymmetry  $\Delta(\omega) \equiv \frac{1}{I_{\mathrm{sd}}}\frac{\partial V_3(\omega)}{\partial B}\Bigr|_{B=0} \propto I_{\mathrm{spin,linear}}$ in the linear response regime, plotted for different fitting ranges of $B$. The results for $I_{\mathrm{sd}} = 5$ nA show that the locations of the extrema of $\Delta(\omega)$ are independent of the fitting range (Fig.~\ref{fig:spin_currents_different_ranges}a), which is consistent with the theoretical prediction that the linear spin signal is proportional to the QPC transconductance. Fig.~\ref{fig:spin_currents_different_ranges}b shows the asymmetry $\Delta(\omega)$ for $I_{\mathrm{sd}} = 32$ nA for different fitting ranges. The magnitude of the asymmetry $\Delta(\omega)$ barely varies with the fitting range. This observation suggests that at low currents the signal is indeed due to a spin accumulation.

\section{The dependence of the non-linear signal on the in-plane magnetic field $B$}
\label{sec:non_linear_in_plane_B}

\begin{figure}[ht]
	\begin{center}
		\includegraphics[scale = 1.15,trim={0cm 0cm 0cm 0cm}]{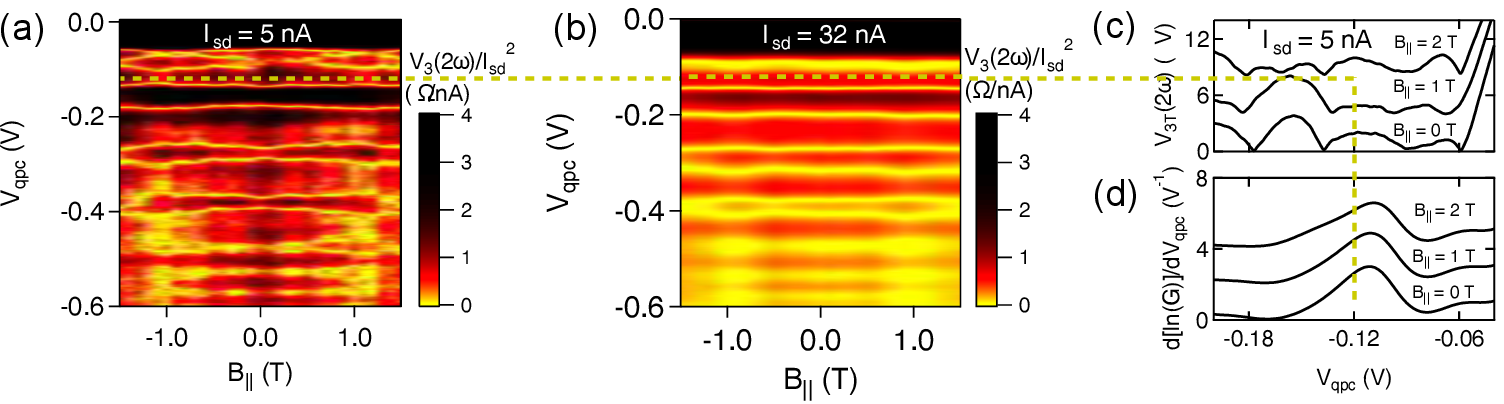}
		\caption{The color map of $V_{3}(2\omega)/I_{\mathrm{sd}}^2$ as a function of the QPC voltage $V_{\mathrm{qpc}}$ and $B$ for (a) $I_{\mathrm{sd}} = 5$ nA and (b) $I_{\mathrm{sd}} = 32$ nA. At 32 nA, the non-linear signal is almost independent of $B$. Panel (c) shows $V_{3}(2\omega)$ as a function of $V_{\mathrm{qpc}}$ at different values of $B$. The lower panel shows the QPC transconductance normalized by the conductance $\frac{\partial{[\ln(G_{\mathrm{qpc}})]}}{\partial V_{\mathrm{qpc}}}$ at various strengths of $B$. (d) Unlike the non-linear signal $V_{3}(2\omega)$ which is fluctuating in $B$, the transconductance is almost unchanged even when $B $ is changed from 0 T to 2 T. Here, the dashed line denotes the second QPC conductance riser.}
		\label{fig:color_maps_of_non_linear_signals}
	\end{center}
\end{figure}

To rule out thermal effects, here we examine the relation between the non-linear spin signal and the in-plane magnetic field, as well as compare it with the relation between the transconductance and the in-plane magnetic field. Concerning the first, Fig.~\ref{fig:color_maps_of_non_linear_signals}a shows the color map of  $V_{3}(2\omega)/I_{\mathrm{sd}}^2$ as a function of the QPC voltage $V_{\mathrm{qpc}}$ and $B$ for $I_{\mathrm{sd}} = 5$ nA and Fig.~\ref{fig:color_maps_of_non_linear_signals}b for $I_{\mathrm{sd}} = 32$ nA. At 5 nA, the non-linear signal varies with $B$. By contrast, at 32 nA, the linear spin signal is suppressed (Fig.~2g and Figs.~4a-d of the main text), and the non-linear signal is now weakly dependent on $B$. The dependence of the non-linear signal on $B$ at 5 nA therefore suggests that in this regime the non-linear signal is spin-related. Concerning the second, we now compare the non-linear signal at 5 nA (Fig.~\ref{fig:color_maps_of_non_linear_signals}c) with the transconductance normalized by the conductance $\frac{\partial{[\ln(G_{\mathrm{qpc}})]}}{\partial V_{\mathrm{qpc}}}$ at various strengths of $B$ (Fig.~\ref{fig:color_maps_of_non_linear_signals}d). We find that while the non-linear signal varies with $B$, the quantity $\frac{\partial{[\ln(G_{\mathrm{qpc}})]}}{\partial V_{\mathrm{qpc}}}$ is almost insensitive to $B$. This suggests that $V_{3}(2\omega)$ is not thermopower, since the latter is approximately proportional to $\frac{\partial{[\ln(G_{\mathrm{qpc}})]}}{\partial V_{\mathrm{qpc}}}$ \cite{Appleyard1998} in contrast to the behavior shown in Fig.~\ref{fig:color_maps_of_non_linear_signals}c.

\section{Comment on the possibility of magneto-thermal effects in our device}

The ac currents we use in the experiments are very small (of order a few nA), which correspond to an input power of $10^{-13}$ W. Therefore, any temperature gradients in the driving channel caused by such amount of power would be only fractions of a Kelvin and would not cause measurable magneto-thermal effects. This means that the current across the QPC is only due to spin-to-charge conversion, where the spin accumulation is generated through the driving current $I_{\mathrm{sd}}$.
 

%